\documentclass[useAMS,usenatbib]{mn2e}

\usepackage{deluxetable}
\usepackage{graphicx}
\usepackage{epsfig}
\usepackage{amssymb,amsmath}
\usepackage{aas_macros}
\usepackage{fixltx2e} % forces figure order to remain fixed
\title[General relativistic disc/jet models of M87]{The size of the jet launching region in M87}
\author[Dexter, McKinney \& Agol]{Jason Dexter$^{1,2}$\thanks{E-mail: 
jdexter@berkeley.edu}, Jonathan C. McKinney$^{3}$ and Eric Agol$^{4}$\\
$^{1}$Department of Physics, University of Washington, Seattle, WA
98195-1560, USA\\
$^{2}$Theoretical Astrophysics Center and Department of Astronomy,
University of California, Berkeley, CA 94720-3411, USA\\
$^{3}$Kavli Institute for Particle Astrophysics and Cosmology, Stanford University, Stanford, CA 94305-4060, USA\\
$^{4}$Department of Astronomy, University of Washington, Box 351580, Seattle, WA 98195, USA}
\begin{document}

\pagerange{\pageref{firstpage}--\pageref{lastpage}} \pubyear{2011}

\maketitle

\label{firstpage}

\begin{abstract}
The supermassive black hole candidate at the center of M87 drives 
an ultra-relativistic jet visible on kiloparsec scales, and its large mass and
relative proximity allow for event horizon scale imaging with
very long baseline interferometry at millimetre wavelengths (mm-VLBI). Recently,
relativistic magneto-hydrodynamic (MHD) simulations of black hole accretion 
flows have proven capable of launching magnetically-dominated jets. We construct 
time-dependent disc/jet models of the innermost portion of the M87
nucleus by performing relativistic radiative transfer calculations from one such
simulation. We identify two types of models, jet-dominated or disc/jet,
that can explain the spectral properties of M87, and use them to make predictions for current and
future mm-VLBI observations. The Gaussian source size for the favored
sky orientation and inclination from observations of the large-scale
jet is $33-44\mu$as ($\simeq4-6$ Schwarzschild radii) on current
mm-VLBI telescopes, very similar to existing observations of Sgr A*. The black hole shadow, direct
evidence for an event horizon, should be visible in future measurements
using baselines between Hawaii and Mexico. Both models exhibit 
variability at millimetre wavelengths with factor of $\simeq 2$
amplitudes on year timescales. For the low inclination of M87, the
\emph{counter-jet} dominates the event horizon scale millimetre
wavelength emission from the jet-forming region.
\end{abstract}


\section{Introduction}

Messier 87 (M87) is a giant elliptical galaxy in the Virgo cluster, known for its galaxy-scale,
ultra-relativistic jet.  
Very long baseline interferometry (VLBI) images at 7mm
\citep{junoretal1999,lyetal2004,walkeretal2008} show extended jet
structure on milli-arcsecond (mas) scales, emanating from an unresolved
bright core. Recent images between $7$ mm and $1.3$ cm \citep{hadaetal2011}
suggest that this core is coincidental with the central black
hole candidate (hereafter M87 refers to the black hole 
rather than the galaxy). M87 is also one of the two largest
black holes on the sky (along with the Galactic centre black hole
candidate, Sgr A*). Its mass is
$\simeq6.4\times10^9 M_\odot$ \citep{gebhardtthomas2009,gebhardtetal2011},
$\sim$1600 times larger than Sgr A*. At a distance of $16$ Mpc, the angular size, $\delta
\theta \propto M/D$, is about 4/5 that of Sgr A*.\footnote{The
  previous black hole mass estimate was a factor of $\simeq 2$ smaller
\citep{marconietal1997}. We use the new black hole mass throughout,
but discuss the effects of the mass on our models in
\S\ref{sec:images}.}

Recent VLBI observations at $1.3$ mm have detected source structure in
Sgr A* on event horizon scales \citep{doeleman2008,fishetal2011}, allowing a direct comparison between
observations and black hole accretion theory
\citep[e.g.,][]{dexteretal2010,brodericketal2011}. These observations
also have the potential to detect the black hole shadow
\citep{bardeen1973,falcke,dexteretal2010}, which would provide the
first direct evidence for an event horizon.

M87 is just as, if not more, promising a mm-VLBI target as Sgr A*. M87 is in
the Northern sky, offering longer mutual visibility with current
telescopes. Its large black hole mass gives a
proportionally longer dynamical time, so that its event horizon spans $\sim$1 light-day. This means
that, unlike in Sgr A*, Earth-aperture synthesis could be used to fill
in the u-v plane, potentially (with additional telescopes) allowing the
creation of an image of the source directly, in contrast with the 
model-dependent
Fourier domain fitting techniques necessary for Sgr A*. Micro-arcsecond ($\mu$as) resolution of mm-VLBI also offers
the possibility of imaging the jet launching region, which would
provide the opportunity to compare directly with physical models of
jet formation in the immediate vicinity of the
black hole. The most recent mm-VLBI campaign observed M87 as
well as Sgr A* \citep{fishetal2011}, but the data are as yet unpublished.

%with constant
%density and magnetic field strength out to some transition radius, where
%both quantities fall off as power laws

The spectral properties of M87 are well known: it is an inverted radio
source with a power law tail extending from the spectral peak in the
millimetre to the optical. Previous semi-analytic work has modeled the low-frequency radio
emission as arising from synchrotron radiation in
either an advection-dominated accretion flow
\citep{reynoldsetal1996,dimatteoetal2003} or a ``truncated'' accretion disc 
\citep{yuan2000,broderickloeb2009}. Both types of models can fit the spectrum. Synthetic jet
images have also been produced as predictions for mm-VLBI from one of
these semi-analytic models \citep{broderickloeb2009}.

Maps of synchrotron emission from numerical simulations of jets have also been compared to
observations of M87
\citep{zakamskaetal2008,graciaetal2009}, but these jets are 
input by hand rather being formed self-consistently from an accretion
flow. Conversely, axisymmetric general relativistic MHD (GRMHD)
simulations of the accretion flow have been used to fit the 
spectrum of the core \citep{moscibrodzkaetal2011,hilburnliang2011}. 

Jet formation has recently become accessible to global, 3D GRMHD
simulations \citep{mckinneyblandford2009}. In their simulation
initialised with a dipolar magnetic field \citep[MBD in][and
here]{dexteretal2010}, an 
ultra-relativistic jet is produced self-consistently from an accretion
disc and propagates stably out to $1000$ M before it interacts with the
simulation boundary.\footnote{Units with $G=c=1$ are used except where
  noted otherwise. In these units, $1$ M for M87 is
  $\simeq 4 \mu$as, $\simeq 10^{15}$ cm, and $\simeq 9$ hours.} We perform
general relativistic radiative transfer via ray tracing to create the first 
time-dependent spectral models of M87 from a GRMHD simulation, and use
the models to make predictions for current and future mm-VLBI
observations. The simulation and the relativistic 
radiative transfer method used to compute observables are described
in \S\ref{sec:methods}, along with the details of the radiative
disc/jet models. Fiducial models for the M87 spectrum are identified
in \S\ref{sec:fiducial-models}, and the resulting millimetre images
and variability are studied in \S\ref{sec:image-morphology} and
\S\ref{sec:variability}. The implications of our results are
discussed in \S\ref{sec:discussion} along with the many uncertainties
in constructing the models. We summarize the results in \S\ref{sec:summary}.

\begin{figure*}
\begin{center}
\begin{tabular}{ll}
\includegraphics[scale=0.12]{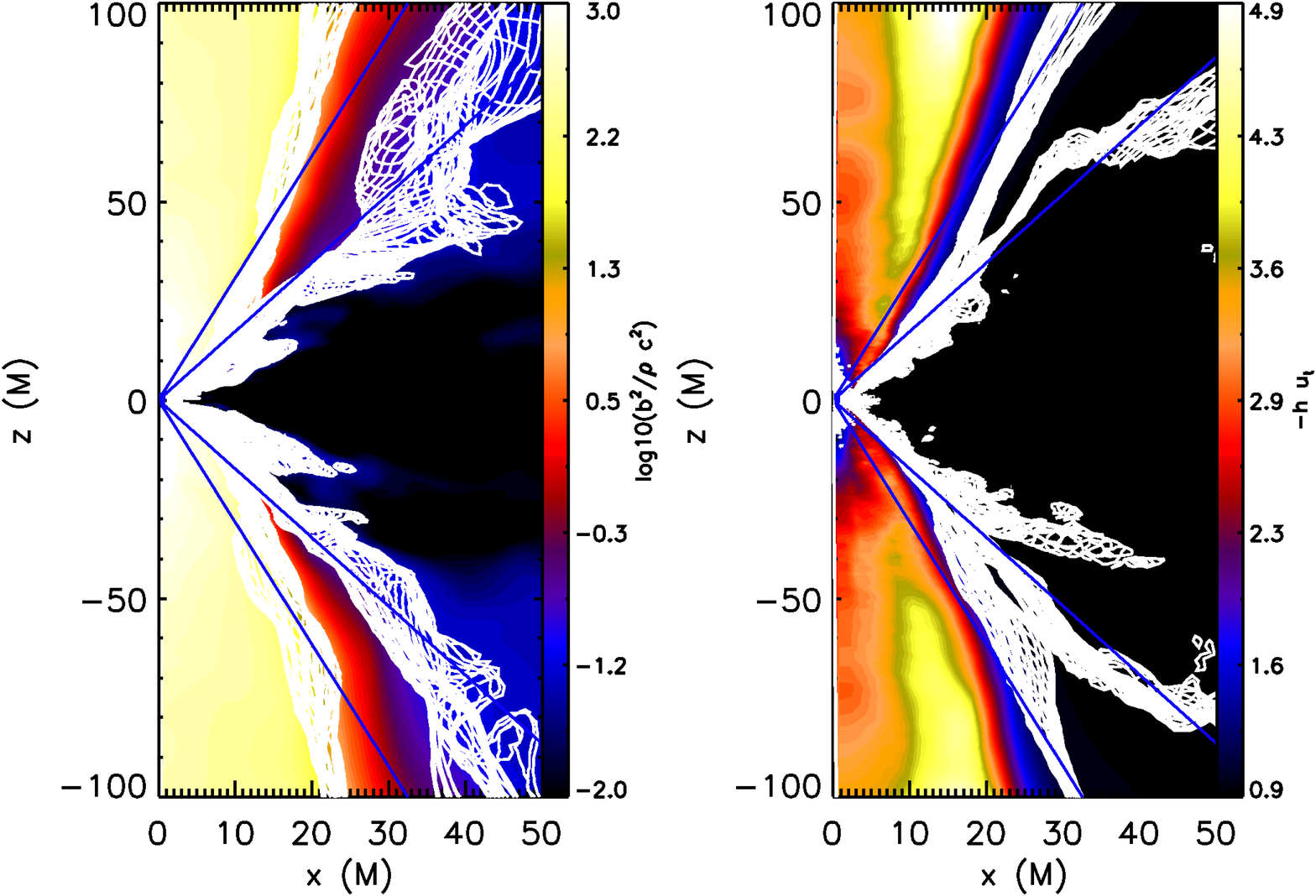}&
\includegraphics[scale=0.12]{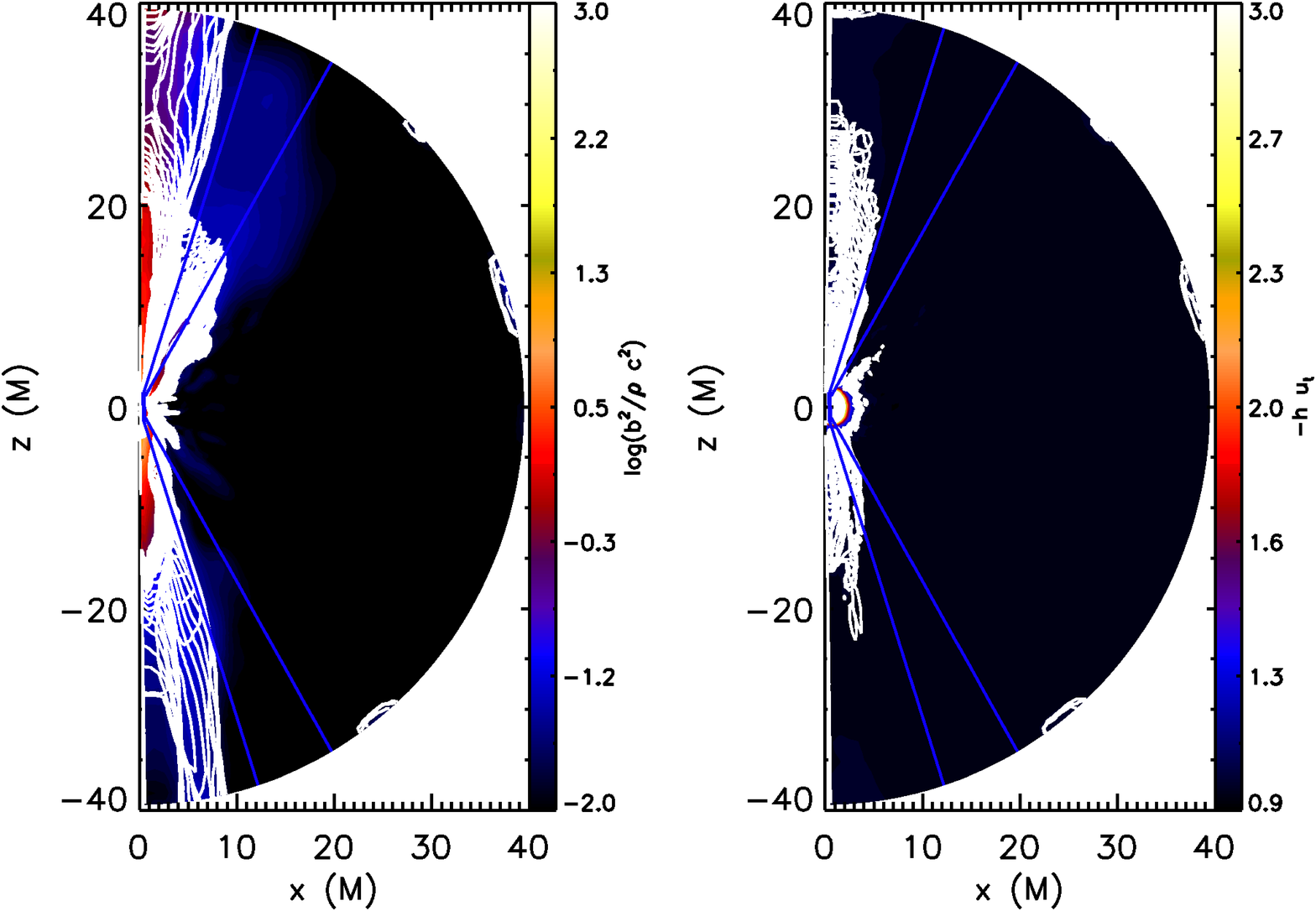}
\end{tabular}
\end{center}
\caption{\label{jetboundary}Ratio of magnetic to rest energy density (first
  and third panels)
  and specific enthalpy measured at infinity (second and fourth panels) for single
  time steps of the MBD (left two panels) and MBQ (right two panels)
  simulations. The azimuthally averaged data are 
  shown in colour, while contours of $b^2/\rho c^2=0.1$, $8\pi$ and $-h
  u_t=1.0, 1.1$ for the full range of azimuths are overplotted. Also
  overplotted in blue are lines of constant polar angle,
  $\theta=18^\circ$, $30^\circ$. For
  M87, $1$ M $\simeq4\mu$as. The jet is clearly identified in MBD, but
  only roughly corresponds to lines of 
  constant $\theta$. There is no persistent jet in the MBQ simulation.}
\end{figure*}

\section{Methods}
\label{sec:methods}
\subsection{Simulation Data}

\citet{mckinneyblandford2009} carried out 3D global GRMHD simulations
starting from a hydrostatic torus in whose angular momentum was aligned with the
black hole spin axis for a spin value $a/M=0.92$. The initial torus
pressure maximum was located at $r=12$ M. The
GRMHD code used was a 3D version of \texttt{HARM}
\citep{gammie2003,noble2006} with 
fourth order interpolation and time-stepping \citep{mckinney2006} as
well as other improvements \citep{mckinney2006ff,tchekhovskoyetal2007}.

This simulation conserved total energy and neglected radiative
cooling.  Energy-conserving simulations convert the energy lost to grid-scale 
numerical magnetic reconnection into heat. This is appropriate for
studying radiatively inefficient sources such as Sgr A* or M87, where
the accretion flow is hot and geometrically thick.\footnote{ \citet{dexter2009} found a radiative
efficiency from energy lost to numerical reconnection of
$\epsilon\simeq 0.1$ for Sgr A* in a non-conservative simulation from
\citet{fragile2007}.} The full azimuthal domain was included, but only marginally resolved with a
resolution of 256$\times$128$\times$32 in ($r$, $\theta$,
$\phi$). The effective resolution was higher than in fully second
order schemes due to the high order scheme used. The coordinates
were regular but warped, with the resolution concentrated toward the mid-plane
at small radius to resolve the disc and towards the pole at large
radius to resolve the jet. The total duration was $3500$ M, with 
roughly constant radial profiles in accretion rate and angular momentum out to $r
\simeq 10$ M by $t \sim 3000$ M. The snapshot used for spectral fitting in 
\S\ref{sec:fiducial-models} is from $t=3000$ M, after an approximate 
quasi-steady state was established in the inner disc. The final
$2000$ M of the simulation is considered when we study time-variable properties of the
radiative models.

\subsection{Ray Tracing}

We performed relativistic radiative transfer on the simulation data
via ray tracing using the code \texttt{grtrans}
\citep{dexter2011}. Starting from an observer's camera, rays are
traced backwards in time toward the black hole assuming they are null
geodesics (geometric optics approximation), using the public code \texttt{geokerr} described in
\citet{dexteragol2009}. In the region where rays intersect the
accretion flow, the radiative transfer equation is solved along the
geodesic \citep{broderick2006} in the form given in
\citet{fuerstwu2004}, which then represents a pixel of the image. This
procedure is repeated for many rays to produce an image, and at many
time steps of the simulation to produce time-dependent images
(movies). Light curves are computed by integrating over the individual
images. Repeating the procedure over observed wavelengths gives a
time-dependent spectrum.

To calculate fluid properties at each point on a ray, the spacetime
coordinates of the geodesic are transformed from Boyer-Lindquist to
the modified Kerr-Schild coordinates used in the simulation
\citep{mckinney2006}. Since the accretion flow is dynamic, light
travel time delays along the geodesic are taken into account. Data
from the sixteen nearest zone centres (eight on the simulation grid
over two time steps) were interpolated to each point on the geodesic.

\subsection{Radiative Modeling}

Computing emission and absorption coefficients requires converting
simulation fluid variables (pressure/internal energy, mass density,
and magnetic
field strength) into an electron distribution function in physical
units. The black hole mass sets the length and time scales, while the
mass of the initial torus provides an independent scale and fixes the
accretion rate. The scalings are such that $n$ and $b^2$ are
proportional to the accretion rate. There is no consensus for the electron distribution 
or geometry responsible for the millimetre emission in M87. The
presence of an extended jet at $7$ mm indicates that the jet is at
least comparable in luminosity to the disc at that wavelength. Since our model
consists of a GRMHD simulation where a jet is produced from accretion
onto a black hole, we also include a disc component. The emission
mechanism is taken to be entirely synchrotron radiation for both
components. The disc electrons are assumed to be thermal,
obeying a relativistic Maxwell (Maxwell--Juttner) distribution. As in
previous models for Sgr A*, we assume a two-temperature flow with a constant ion-electron temperature
ratio which is left as a free parameter
\citep{goldston2005,moscibrodzka2009,dexteretal2010}. With this
parameter, the electron
temperature is computed from the gas pressure in the simulation using
the ideal gas law. We use the
approximate angle-dependent, unpolarized emissivity from
\citet{leungetal2011}. Previous models of M87 have also 
included non-thermal disc emission, which could be important for
explaining the radio spectrum \citep{broderickloeb2009}. 

\begin{figure*}
\begin{center}
\begin{tabular}{c}
\includegraphics[scale=0.8]{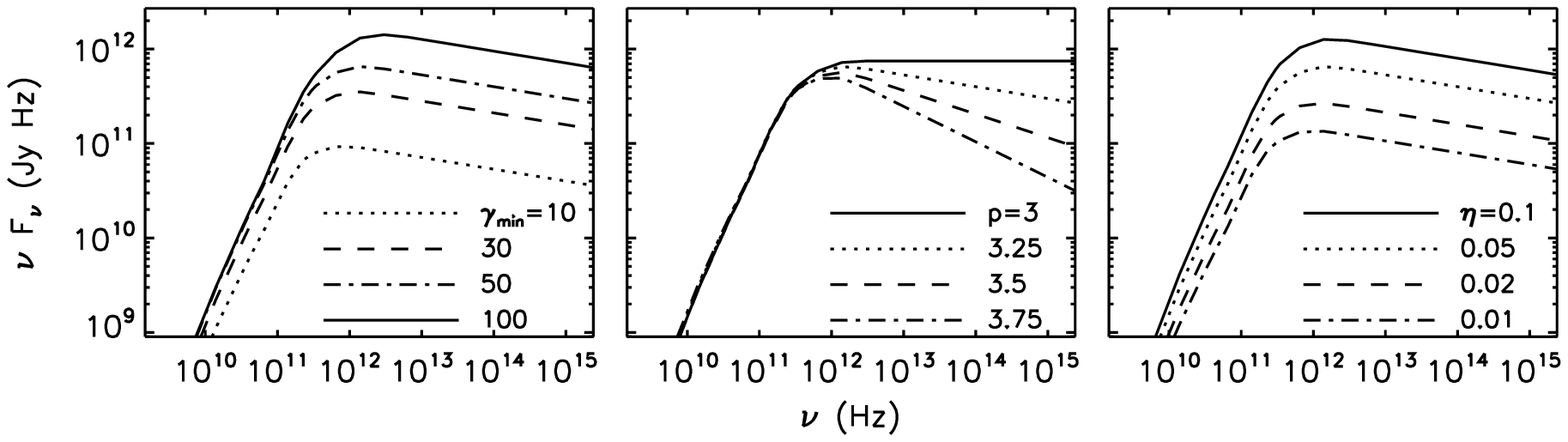}\\
\includegraphics[scale=0.8]{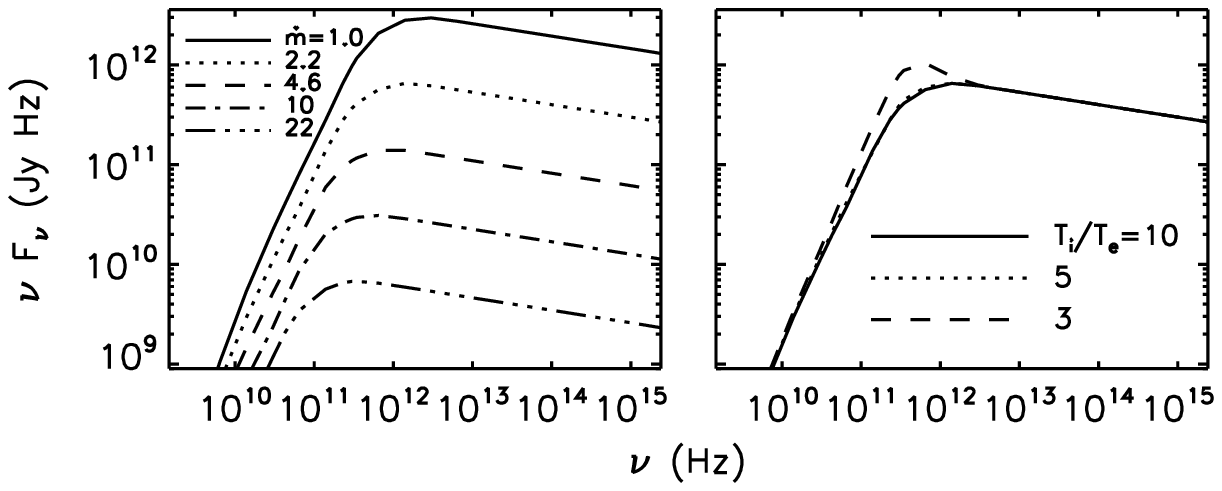}
\end{tabular}
\end{center}
\caption{\label{samplespecparam}Sample disc/jet spectra varying each parameter
  independently: $\gamma_{\text{min}}$ (top left), $p$ (top middle), $\eta$
  (top right), $\dot{m}$ (bottom left, in units of $10^{-5}$), and $T_i/T_e$ (bottom
  right). The default parameter values are $\gamma_{\text{min}}=50$,
  $p=3.25$, $\eta=0.05$, $\dot{m}=10\times10^{-5}$ and $T_i/T_e=10$.}
\end{figure*}

\subsection{Jet Emission}

We assume that the jet emission is from non-thermal electrons, whose particle distribution
is a power law in electron energy (Lorentz factor) with a constant index
$p$ between low- and high-energy cutoffs,
$\gamma_{\text{min},\text{max}}$. The unpolarized 
synchrotron emission and absorption coefficients, $j_\nu$ and $\alpha_\nu$, for this
distribution in cgs units are \citep{leggwestfold1968,yuanquataert2003,dexter2011}, 

\begin{align}\label{plemis}
j_\nu = \frac{ne^2(p-1)\nu_c}{2\sqrt{3}c
  (\gamma_{\text{min}}^{1-p}-\gamma_{\text{max}}^{1-p})}\left(\frac{\nu}{\nu_c}\right)^{-\frac{p-1}{2}}&\nonumber\\
\times\left[G(x_{\text{min}})-G(x_{\text{max}})\right],&\\
\alpha_\nu =
\frac{ne^2(p-1)(p+2)}{4\sqrt{3}mc\nu_c(\gamma_{\text{min}}^{1-p}-\gamma_{\text{max}}^{1-p})}\left(\frac{\nu}{\nu_c}\right)^{-\frac{p}{2}-2}&\nonumber\\
\times\left[Ga(x_{\text{min}})-Ga(x_{\text{max}})\right],&
\end{align}

\noindent with $\nu_c=\frac{3eB\sin{\theta}}{4\pi m c}$,
$x_{\text{min},\text{max}} = \nu/(\gamma_{\text{min},\text{max}}^2
\nu_c)$, $\nu$ is the emitted frequency, $n$ is the particle density,
$B$ is the magnetic field strength, $\theta$ is the angle between the
photon wave-vector and the magnetic field, and the power law synchrotron
integrals are, 

\begin{align}
G(x)&=\int_x^\infty dz z^{\frac{p-3}{2}} F(z),\\
Ga(x)&=\int_x^\infty dz z^{\frac{p}{2}-1} F(z),
\end{align}

\noindent and where,

\begin{equation}
F(x)=x \int_x^\infty dy K_{5/3} (y),
\end{equation}

\noindent is the unpolarized synchrotron function and $K_\alpha (y)$
is a modified Bessel function. The functions
$G(x)$ and $Ga(x)$ are tabulated for the desired
values of $p$ and interpolated for the emissivity
calculation. Evaluating the double integrals for constructing the
interpolation tables can be
sped up significantly using \citep{westfold1959},

\begin{align}
\int_x^\infty d\xi \xi^{s-1} \int_\xi^\infty dy& K_\alpha(y)=\frac{\alpha+s}{s}\int_x^\infty d\xi \xi^{s-1} K_\alpha(\xi)\nonumber\\
&-\frac{x^s}{s}\left[\int_x^\infty dy K_{\alpha+1}(y)-K_\alpha(x)\right]
\end{align}

\noindent to convert them to sums of single integrals. This form for the absorption
coefficient is original, derived in the same fashion as for the
emissivity in previous work. Both coefficients reduce to the standard
approximate forms when $x_{\text{min}}=0$ and $x_{\text{max}} \rightarrow \infty$.

This form of the emissivity is more accurate for M87 than approximate forms commonly found in
the literature that assume a frequency far from those corresponding to
the cutoff Lorentz factors. In M87, the frequency corresponding to the
low-energy Lorentz factor cutoff, for 
$\gamma_{\text{min}}=10-100$, is $\nu_0 \simeq
10^{10-11}$Hz, close to the
frequencies of interest for mm-VLBI ($230$ GHz and $345$ GHz). Taking the low-frequency cutoff
into account broadens the spectrum and smoothes the turnover from
optically thick to thin. 

The magnetic field strength everywhere is taken directly from the
simulation. For the jet emission, we need to calculate a
non-thermal particle density. In magnetically-dominated regions such
as the jet, the particle density and internal energy from the
simulations are highly inaccurate due to the artificially enforced
floor values used for numerical stability. Instead of using these
compromised values, we scale the internal energy to the
magnetic energy with a constant of proportionality, $\eta$
 \citep[cf.][]{broderickmckinney2010}: 

\begin{equation}
\label{eq:uproptob}
u_{\text{nth}} = \eta \frac{b^2}{8\pi},
\end{equation}

\noindent where $u_{\text{nth}}$ is the non-thermal internal energy density and $b$ is
the magnetic field strength in cgs units. Then the particle density, $n_{\text{nth}}$, is
taken from,

\begin{equation}
\label{eq:nnth}
n_{\text{nth}}=\eta \frac{b^2}{8\pi} \frac{p-2}{p-1} (m_e c^2 \gamma_{\text{min}})^{-1},
\end{equation}

\noindent where $m_e$ is the electron mass and which implicitly assumes that all of the internal energy is
in electrons, or equivalently that the thermal
energy in all particles is negligible and that the positive charge
carriers are thermal. 

For self-consistency, the rest energy of these non-thermal particles should still be less
than the magnetic energy density. This leads to the condition, 

\begin{equation}
  \eta \lesssim \frac{m_e}{m_{e,p}} \frac{p-1}{p-2} 8\pi \gamma_{\text{min}},
\end{equation}

\noindent where $m_{e,p}$ correspond to leptons (baryons) producing
the jet. The strictest condition on $\eta$ is found by assuming a
baryonic jet, in which case (for $p=3-3.5$),

\begin{equation}\label{eq:eta}
\eta \lesssim .25 \frac{\gamma_{\text{min}}}{10}.
\end{equation}

\noindent This inequality is satisfied in all our models as the
maximum $\eta$ considered is $0.1$.

\begin{figure*}
\includegraphics[scale=0.7]{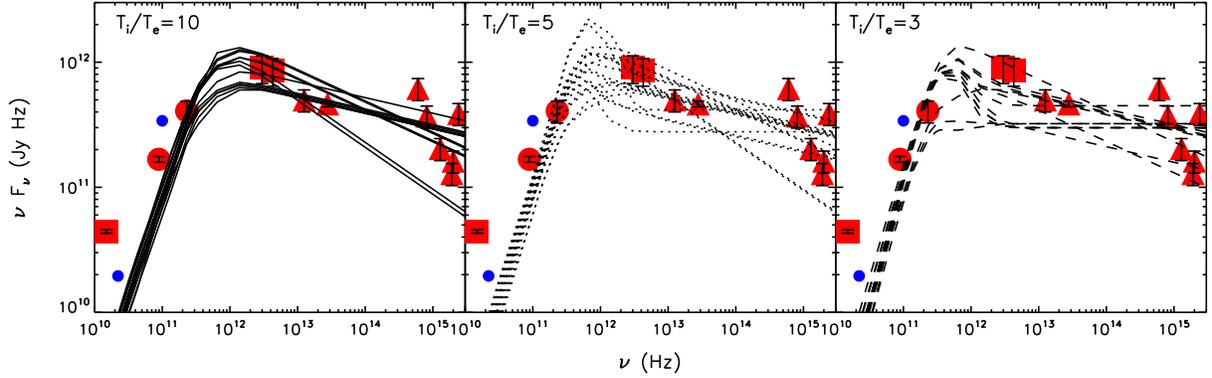}
\caption{\label{m87spectra}Viable spectra with $T_i/T_e=10$ (left),
  $5$, (middle), $3$ (right). The large red symbols are
the observational data used as either upper limits (squares), data
points (circles) or waveband-averaged data points (triangles). The
small blue circles are the flux in the unresolved radio core of VLBI
observations.}
\end{figure*}

\subsection{Disc/Jet Boundary}

The GRMHD simulation consists of a solution for a single
component fluid, neglecting electrons and dynamical effects on
the local particle distribution. Defining a disc/jet solution then requires
choosing a condition for the boundary between the two
components. There are a few possibilities.

The general structure of numerical simulations of black hole accretion
flows consists of a dense,
thick disc with scale height $H/R \sim 0.2-0.3$ centered on the equatorial
plane, surrounded by a tenuous wind and then a polar jet. One
straightforward method is to define the disc/jet boundary at a
particular polar angle ($\simeq 30^\circ$).

The outflows in GRMHD simulations arise in
magnetically-dominated regions
\citep{devilliersetal2005,mckinney2006}. The second method is then to
define the jet as anywhere that $b^2/\rho c^2 > f$, where $f\simeq0.1-8\pi$
depending how much of the ``disc wind'' is to be included in the jet.

Finally, the disc/jet components can be separated as the boundary
between inflow/outflow. In this case, unbound fluid elements
constitute the jet ($-h u_t > 1$, where $h$ is the specific 
enthalpy and $u_t$ is the time-component of the covariant fluid
four-velocity, \citealt{devilliersetal2005}). Any combination of
these criteria can also be used. A comparison of
the three criteria for an azimuthally-averaged time step of the
simulation is shown in Figure \ref{jetboundary}, and as expected
they all lead to roughly the same definition of the jet for MBD. The
inflow/outflow and magnetically-dominated criteria are in particularly
good agreement, while the constant slice in polar angle tends to cut out
a significant amount of the jet base crucial for mm-VLBI images. We also show
the same criteria applied to the simulation from
\citet{mckinneyblandford2009} initialised with a large-scale
quadrupolar magnetic field \citep[MBQ in][]{dexteretal2010}, where no persistent 
jet forms. Little of the polar region in this simulation would qualify
as a jet by our criteria that fluid be magnetically-dominated or
unbound.

In the fiducial models discussed below, we use the condition $b^2/\rho
c^2 > 1$, but our results are insensitive to the exact criterion used to
separate the disc and jet components.

\subsection{Model Parameters}

These are the required elements to define a radiative model of
M87. Unlike in Sgr A*, the inclination of M87 is fairly well
constrained from observed superluminal motion in the jet
\citep{heinzbegelman1997}, implying $i \lesssim 30^\circ$. We fix
$i=25^\circ$ in this work, but discuss the effect of varying it in
\S~\ref{sec:images}. The black hole spin is also fixed at 
$a/M=0.92$ since we only consider models from the MBD simulation.

Each radiative model of M87 then has the following parameters: the
ion-electron temperature ratio $T_i/T_e$ for the disc component, the
accretion rate $\dot{M}$ and the disc/jet boundary
selection criterion for both components, and
$\gamma_{\text{min},\text{max}}$, $p$, and $\eta$ for the jet
component. We fix $\gamma_{\text{max}}=10^5$ throughout
\citep[see][]{broderickloeb2009}, leaving five free parameters. This
is many more free parameters than in prior radiative models of Sgr A* from
GRMHD simulations 
\citep{moscibrodzka2009,dexter2009,dexteretal2010,shcherbakovetal2011}. 
Sample spectra showing the effects of independently
varying the parameters are shown in Figure \ref{samplespecparam}. The
ion-electron temperature ratio $T_i/T_e$ (bottom right panel) determines the relative disc
contribution. At $T_i/T_e=10$, the disc emission is negligible at all
wavelengths while at $T_i/T_e=3$, the best-fitting ratio for this
simulation for Sgr A* \citep{dexteretal2010}, the jet portion dominates except in the
millimetre, where thermal disc emission produces a sub-millimetre
bump. The normalization and peak frequency are affected by both 
$\gamma_{\text{min}}$ (for the jet portion, upper left panel) and $\dot{m}$
(bottom left panel). The fraction of magnetic energy converted into
non-thermal jet particles, $\eta$, changes the normalization of the
jet spectrum (top right panel). Finally, $p$ fixes the spectral
slope between the millimetre and IR/optical emission.

\begin{table}
\caption{Fiducial Model Parameters \label{m87_fiducial}}
\begin{small}
\begin{center}
\begin{tabular}{lccccc}
        \tableline
	\tableline
Model & $T_i/T_e$ & $\dot{m} (10^{-4})$ & $\gamma_{\text{min}}$ & $p$ & $\eta$\\
        \tableline
DJ1 & $3$ & $1$ & $50$ & $3.25$ & $0.05$ \\
% & $5$ & $2.2$ & $30$ & $3.25$ & $0.02$ \\
J2 & $10$ & $1$ & $50$ & $3.50$ & $0.10$\\
	\tableline
\end{tabular}
\end{center}
\end{small}
\end{table}

\section{Fiducial Models}
\label{sec:fiducial-models}
To identify viable models, we compute spectra from a single
time step of the MBD simulation data over a grid spanning reasonable values of
the various parameters: $T_i/T_e=(3, 5, 10)$, $\dot{m}=(1,
2.2, 4.6, 10, 22, 46, 100) \times 10^{-5}$, $p=(3, 3.25, 3.5, 3.75)$,
$\gamma_{\text{min}}=(10, 30, 50, 100, 1000)$ and $\eta=(0.01, 0.02, 0.05,
0.10)$. The relative accretion rate is defined as $\dot{m} \equiv
\dot{M}c^2/L_{\text{edd}}$, where $L_{\text{edd}} \simeq
8\times10^{47} (M/6.4\times10^9 M_\odot)\hspace{2pt}\text{erg}\hspace{2pt}\text{s}^{-1}$ is the Eddington
luminosity. Although we only use a snapshot of the simulation, the shape of the spectra
as well as the morphology of the millimetre wavelength images (see
\S\ref{sec:image-morphology}) do not change significantly over the
simulation, even as the total flux varies by a factor of a few (see
\S\ref{sec:variability}). In \S\ref{sec:predictions-mm-vlbi}, we discuss
the effect of time variability on the predicted emission region
sizes for mm-VLBI.

The resulting spectra are fit to multi-wavelength observations. We fit
to average values in the optical \citep{sparksetal1996} and near infrared
\citep{perlmanetal2001,perlmanetal2007} and to the measured values at $3.3$ mm and $1.3$ mm
\citep{tanetal2008}. The far infrared measurements are
treated as upper limits due to their possible contamination by dust in
the host galaxy \citep{perlmanetal2007}. The radio data are also treated
as upper limits, since as in Sgr A* the spatial extent of the 
simulation is too limited to model large-scale emission. The uncertainties are taken as $30\%$ in all
cases irrespective of measurement errors, since we are interested in
finding qualitatively reasonable spectra rather than quantitatively
constraining parameters. 

Examples of the many feasible models from the grid of spectra are shown in Figure \ref{m87spectra}. The lines are all
spectra for which $\chi^2 < 0.5$, split up by ion-electron temperature
ratio. The $\chi^2$ values are low because of our 
artificial inflation of the error bars, and the number of
observational constraints is equal to the number of free parameters in
our model. For both of these reasons we make no attempt to
quantify a goodness of fit for the model spectra.

These disc/jet models are much different from previous semi-analytic 
models of M87 \citep{yuan2000,broderickloeb2009}. The jet and disc
emission both peak in the millimetre, unlike truncated disc models
where the disc emission peaks in the radio. The peak frequency of the
disc spectrum and its normalization are fixed by $T_i/T_e$ and
$\dot{m}$. Both parameters scale the
normalization of the spectrum with its peak frequency, so that it is not possible to produce
the observed radio emission from the GRMHD accretion flow. It could,
however, be produced by thermal or non-thermal disc electrons at large
radius outside of the simulation volume. Then the truncated disc would
be most closely related to our models with $T_i/T_e=10$, where the
disc component contributes negligibly. The disc components of the model spectra with
$T_i/T_e=3$ are similar to the time-averaged spectra from axisymmetric GRMHD simulations
\citep{moscibrodzkaetal2011,hilburnliang2011}, except that we neglect
the effects of Compton scattering. 

%The radiative transfer calculation
%in \citet{hilburnliang2011} is also non-relativistic, which is
%inappropriate for the event horizon scale emission at millimetre
%wavelengths. 

The jet spectrum has the additional degrees of freedom $\eta$ and
$\gamma_{\text{min}}$, which allow the possibility of a spectral peak in the
radio. However, this is seemingly in conflict with $7$ mm VLBI
observations, which find extended mas scale jet emission \citep{junoretal1999}. In our jet models and those from
\citet{broderickloeb2009}, the emission is extended when
optically thick to synchrotron self-absorption, and the existing VLBI
observations suggest that the jet spectrum is still rising in the
millimetre. Moving the spectral peak of the jet component also
requires extreme values of the relevant parameters, with 
$\dot{m} \gtrsim 10^{-3}$, $\eta \sim 1$ and $\gamma_{\text{min}} \sim
1$. For both reasons, we do not pursue these models further.

\begin{figure}
\includegraphics[scale=0.65]{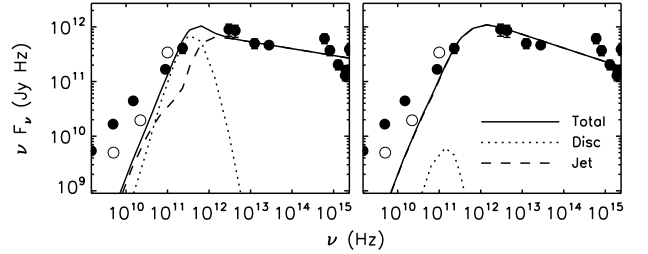}
\caption{\label{m87specfid}Total (solid), disc (dotted) and jet
  (dashed) spectra from the two fiducial models along with the
  observational data points. The error bars shown are from the
  measurements, whereas $30\%$ uncertainties were
  used for the fitting. In the model with $T_i/T_e=3$ (left
  panel), the thermal disc electrons produce the millimetre emission. For small electron
  temperatures (right panel), the entire spectrum is from the
  non-thermal jet component.}
\end{figure}

\subsection{Viable Parameter Ranges}

We select the best-fitting spectra to the chosen data with $T_i=3$, $10$
as fiducial models representative of the set of viable possibilities. This parameter
fixes the relative contribution of the disc. The parameters of the two
fiducial models (DJ1 and J2) are listed in Table \ref{m87_fiducial}, and
their spectra are plotted in Figure \ref{m87specfid}, showing the
total spectra as well as their separate jet and disc components. The
solid points are the total observed flux, while the open circles show
only the flux from the unresolved radio core of VLBI observations
\citep{paulinytothetal1981,spencerjunor1986,baathetal1992}. This is a
better comparison for the extremely small scale emission seen in our
models, and gives reasonable agreement for frequencies $\gtrsim
10^{10} \text{Hz}$.

Typical parameter values producing the millimetre emission in
  the fiducial models are $n
\lesssim 10^7 \hspace{2pt} \text{cm}^{-3}$, $b \simeq 10$ G, and $T_e
\simeq 2\times10^{10}$ K. It is important to note that the observed
emission depends on the electron temperature itself, despite the
common parametrization in terms of $T_i/T_e$. Simulations with hotter
ions would favor similar values of $T_e$, but much larger values of
$T_i/T_e$. Typical non-thermal jet particle densities are
$n_{\text{nth}} \lesssim 10^4 \hspace{2pt} \text{cm}^{-3}$.

\subsection{Jet Only or Disc/Jet}

With $T_i/T_e=10$, the disc emission is negligible at all
wavelengths. The jet spectrum peaks in the millimetre, with a power law tail
extending to the IR/optical. Models with $T_i/T_e=3$ are still
jet-dominated at low and high frequencies, but the thermal emission
leads to a sub-millimetre bump. This not only leads to significant disc
emission at frequencies of interest for mm-VLBI, but the thermal
absorption can also attenuate the jet emission, which for the
inclination of M87 is mostly from the \emph{counter-jet} (see 
\S\ref{sec:image-morphology}). The spectrum alone cannot distinguish
between the jet or disc electrons producing the millimetre emission.

\begin{figure*}
\includegraphics[scale=1.0]{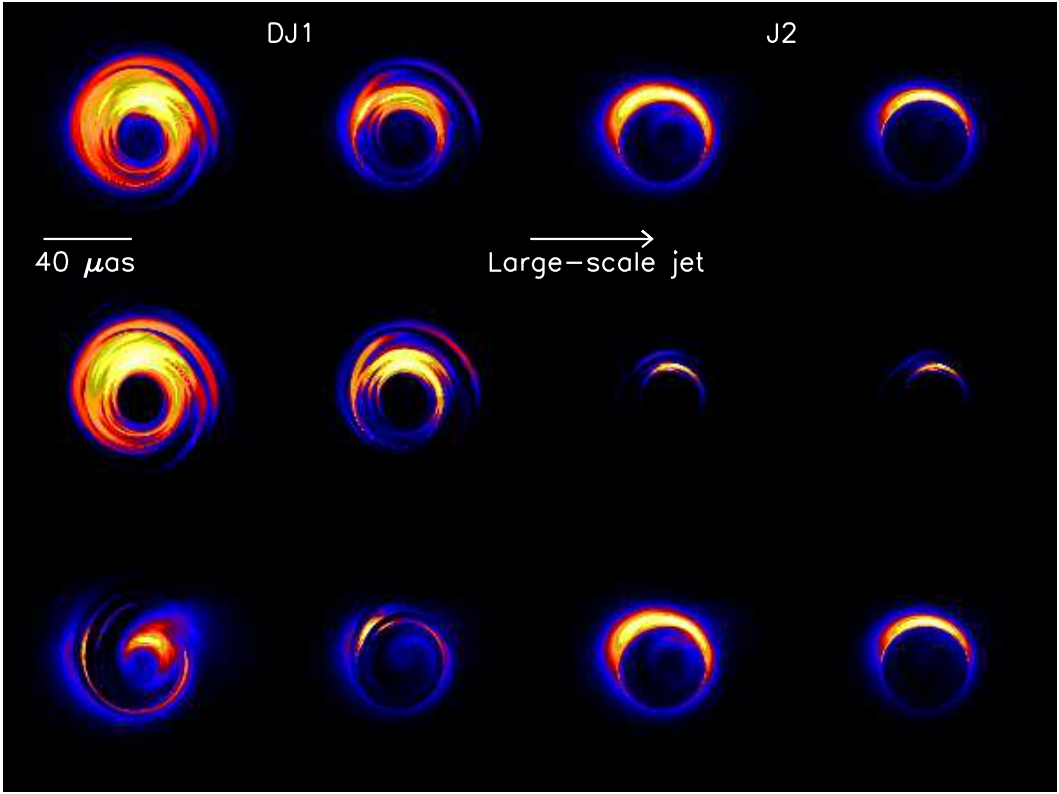}
\caption{\label{m87imgs}Images from total (top row), disc (middle row)
  and jet (bottom row) components for the two fiducial models (left
  two columns and right two columns) at $1.3$ mm (first and third
  columns) and $0.87$ mm (second and fourth columns). The colours are
  scaled linearly from blue to red to yellow to white, with a dynamic
  range of 60. The panel size is $100\times100\mu$as. The images are
  taken from the same time step used for the spectra in Figure
  \ref{m87specfid}, and have been rotated $90^\circ$ to roughly align
  with the position angle of the large-scale jet at $7$ mm. Both
  fiducial model images are crescents from the combined effects of
  light bending and Doppler beaming. The disc component of DJ1 is
  similar to previous models of Sgr A*. The jet component in J2 is
  dominated by the counter-jet, while in DJ1 the jet component comes
  from the forward jet since the counter-jet emission is absorbed by
  the disc.}
\end{figure*}

\section{Image Morphology}
\label{sec:image-morphology}
Images of the fiducial models as well as their jet and disc
components from the time step of the MBD simulation used for spectral 
fitting are shown in Figure \ref{m87imgs} at $1.3$ mm and $0.87$ mm, 
the two wavelengths of interest for mm-VLBI. The orientation angle of
the M87 black hole spin axis projected on the plane of the sky can be
reasonably assumed to align with the orientation of the $7$ mm jet
structure. This assumes that the jet is launched along the spin axis,
and that the jet remains coherent on parsec scales. The images in
Figure \ref{m87imgs} have all been rotated to this favored orientation. 

Images of M87 dominated by thermal particles in the accretion flow
(DJ1) are nearly identical to those of Sgr A* \citep[cf. Figure 11
of][]{dexteretal2010}. Doppler beaming from the Keplerian velocity
profile is significant even at the low expected inclination of M87
($i=25^\circ$ here), but weaker than for preferred inclinations of Sgr
A* ($\simeq 60^\circ$). The image is a crescent from the combined
effects of beaming, light bending and gravitational lensing. The
emission region is in the inner portion of the accretion flow ($r \sim
5M$) near the mid-plane
\citep[see][]{moscibrodzka2009,dexteretal2010}. 

%, and these lines of sight have longer path lengths and less
%severe combined Doppler and gravitational redshifts. 

The jet emission arises from near the pole at very small radius ($r
\sim 2-4M$), which falls mostly in the
``shadow'' region of the image, connected with an observer at infinity
by photons that intersect the black hole. The portions outside the shadow for
lines of sight at the low inclinations considered here ($i \lesssim
40^\circ$) are in the \emph{counter-jet} rather than the forward
jet. Inside the shadow, a small fraction of the emitted photons reach
a distant observer, strongly suppressing the brightness of the forward
jet. When the disc emission is negligible (J2),
the lines of sight to the counter-jet are optically thin and the
counter-jet dominates 
the jet image. The vertical velocity components are
small at these small radii, so that the Doppler beaming is dominated
by the azimuthal velocity component. This leads to the asymmetry in the
jet component of the image. The jet image is still a crescent,
although its intensity drops off sharply due to the strong radial dependence of the
jet emissivity ($j_\nu \sim n_{\text{nth}}B^2 \sim r^{-4}$ with $B \sim
r^{-1}$ near the spectral peak). When the disc emission is
significant, lines of sight to the  
counter-jet become optically thick, and the jet emission is dominated
by the forward jet contribution (lower left two panels of Figure
\ref{m87imgs}). 

\begin{figure*}
\includegraphics[scale=1.0]{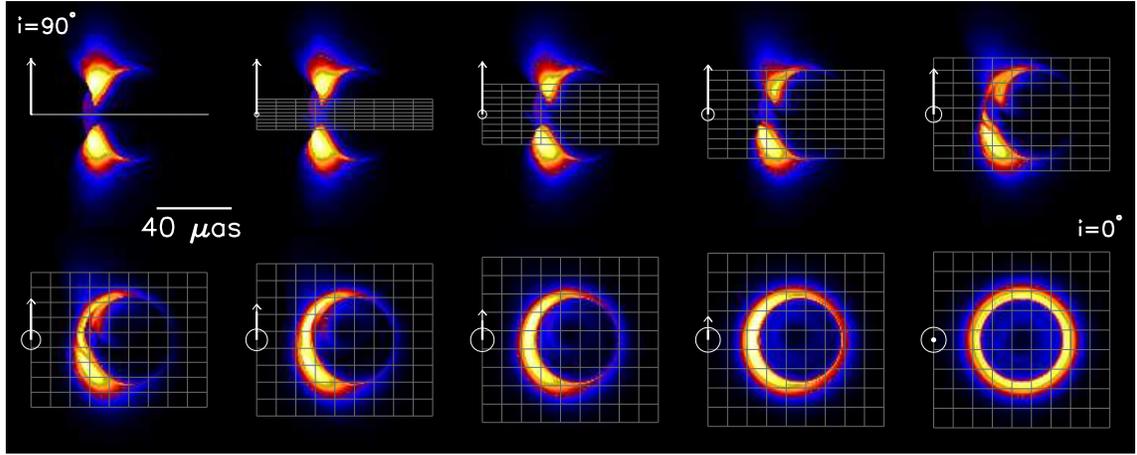}
\caption{\label{jetimages}Jet images at observer inclinations ranging
  from edge-on (top left panel) to face-on (bottom right panel). The colours are
  scaled linearly from blue to red to yellow to white, with a dynamic
  range of 60. The panel size is $100\times100\mu$as. The arrow shows
  the orientation of the black hole spin axis, and the grid is 
  the xy-plane. For edge-on viewing there is strong asymmetry from
  Doppler boosting and no distinction between the jet/counter-jet, 
but as the inclination decreases the image becomes circularly 
symmetric and the counter-jet becomes more prominent as more of the 
forward jet emission is captured by the black hole.}
\end{figure*}

These are the first images of a jet launching region from a
simulation. Images of the same time step are plotted in
Figure \ref{jetimages} for inclinations of $90^\circ$ (edge-on, top
left panel) to $0^\circ$ (face-on, bottom right panel) in even steps
of $\cos{i}$, where $i$ is the inclination angle. For edge-on viewing,
there is no distinction between jet 
and counter-jet and the image is split into two bright lobes close to
the black hole. The image is highly asymmetric due to strong Doppler
beaming from helical motion in the jet base. Moving to lower
inclination, more and more of the 
forward jet falls into the shadow region where the path lengths are
shorter and less of the orbital velocity is aligned with the line of
sight. This causes the brightening of the counter-jet relative to the
forward jet. At small inclinations, the forward jet is barely visible
and the image is dominated by the counter-jet. This is the opposite of
the behavior seen in extended jet emission where the material is
ultra-relativistic and the radial velocity dominates. In that case,
the forward- (counter-) jet is strongly Doppler boosted towards 
(away from) the observer at small inclinations. The 
counter-jet in M87 isn't visible even on mas scales at $7$ mm. These
jet images are also much different from those in 
\citet{broderickloeb2009}, where the jet is launched farther from
the black hole ($r \sim 10M$) than is found from the simulation assuming that
$n_{\text{nth}} \propto b^2$. When the jet is launched farther from
the black hole, the forward jet dominates the images at $1.3$ mm. For
this reason, the jet images discussed here are sensitive to the
uncertain mass loading in ultra-relativistic jets. Jet particle
density profiles for M87 found from pair production calculations are
qualitatively similar to those used here \citep{moscibrodzkaetal2011}.

In both the disc and jet models the circular photon orbit produces a
black hole shadow at $1.3$ mm. At $0.87$ mm, both images
are more compact. The jet component in DJ1 is less significant, and the J2
image is essentially just a single crescent from Doppler beaming of
the jet material with a velocity component along the line of sight.

%\footnote{The favored orientation
%  is ambiguous due to our ignorance of the jet helicity. However, the
%  observed visibilities are the Fourier transform of the real-valued
%  brightness distribution so that $V^*(-u,-v)=V(u,v)$ and their
%  amplitudes, considered here, are identical under this
%  transformation.}

\subsection{Predictions for mm-VLBI}\label{sec:predictions-mm-vlbi}

\begin{figure*}
\includegraphics[scale=1.0]{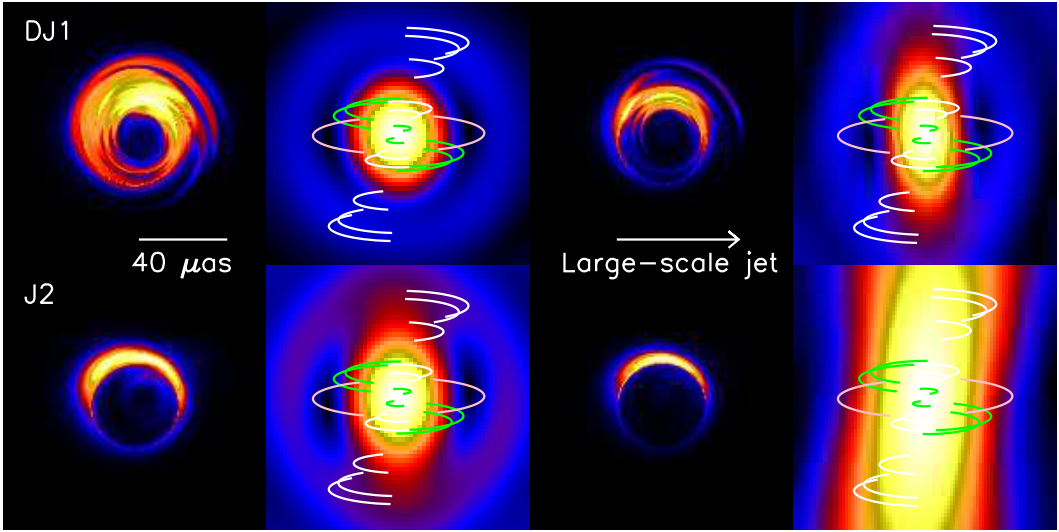}
\caption{\label{m87vis}Images (first and third columns) and
  corresponding visibility amplitudes (second and fourth columns) from
the two fiducial models (rows) at $1.3$ mm (first two columns) and
$0.87$ mm (right two columns). The colours are
  scaled linearly from blue to red to yellow to white, with a dynamic
  range of 60. The panel size is $100\times100\mu$as. The baseline
  orientations available to current 
(green) and near future (white and pink) telescopes are overplotted as
lines. At this orientation, set by the direction of the $7$ mm jet, the
black hole shadow is accessible to future observations on a
Hawaii-Mexico baseline.}
\end{figure*}

\begin{figure}
\begin{center}
\begin{tabular}{l}
\includegraphics[scale=0.7]{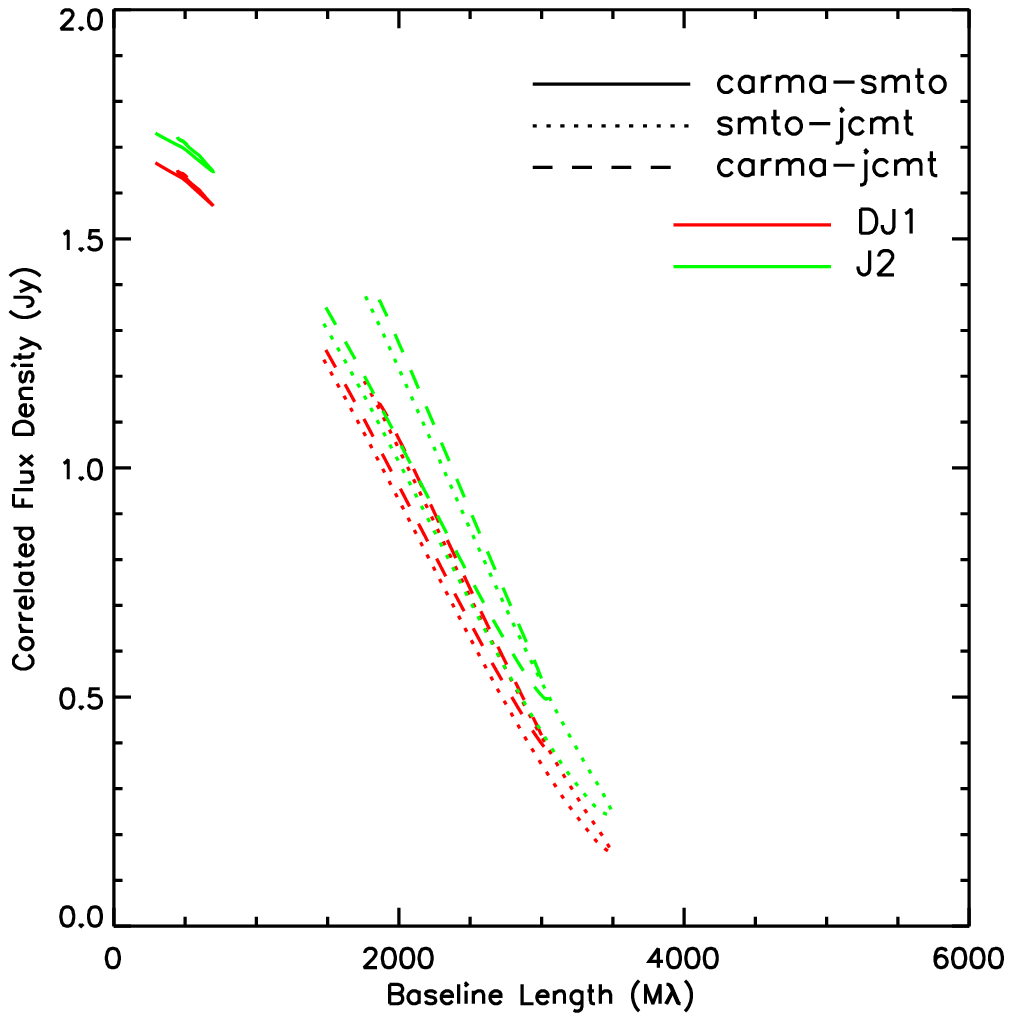}\\
\includegraphics[scale=0.7]{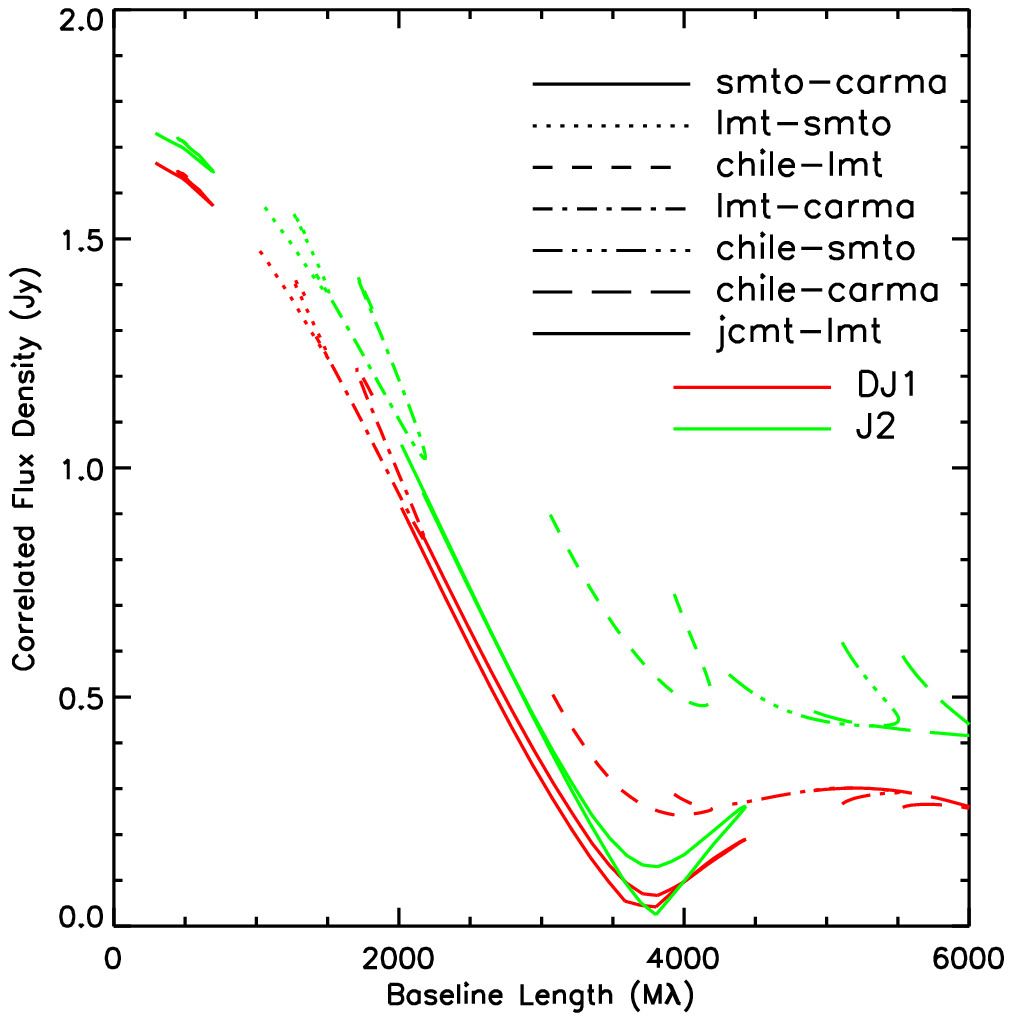}
\end{tabular}
\end{center}
\caption{\label{m87interpvis}Model visibility amplitude vs. baseline length
for current (top) and near future (bottom)
telescopes. The visibility amplitude falls off monotonically with
baseline length for current measurements and can be fit reasonably
well with a circular Gaussian model. The black hole shadow appears as
a local minimum in the visibility profile, and is accessible to a future
baseline between Hawaii and Mexico.}
\end{figure}

The first mm-VLBI observations of M87 were conducted recently
\citep{fishetal2011} using telescopes in Arizona, California, and
Hawaii,\footnote{Submillimeter Telescope Observatory, SMTO; Combined
  Array for Research in Millimeter Astronomy, CARMA; and James Clerk
  Maxwell Telescope, JCMT.} but the results are not yet available. We 
therefore make predictions for both current and future telescopes,
assuming a geometry (inclination and orientation) based on larger
scale jet observations. With sufficient baseline coverage,
sensitivity and observing time it may be possible to construct
observational images in the future. However, for now the small number
of available telescopes requires that models be fit to observations in the u-v plane. 
We Fourier transform the images to visibilities, whose 
amplitudes are shown along with the images for both
mm-VLBI wavelengths in Figure \ref{m87vis}. Possible locations of
current measurements are shown as the green lines, and the visibility
amplitudes are interpolated to those locations and plotted against
baseline length in the top panel of Figure \ref{m87interpvis}. In both fiducial models the current telescopes are
at an orientation where the visibility amplitude decreases
monotonically with baseline length. Fit with a symmetric Gaussian
model, the source FWHM size is 43 (36) $\mu$as for this single time
step of model DJ1 (J2). 

However, the models are time variable and sensitive to the sky
orientation. To make a more robust prediction, these effects are taken
into account by fitting symmetric Gaussian models to visibilities from
120 images spaced evenly over the last $\simeq2000$ M of simulation time ($\simeq 2$ years for M87) with sky
orientations of $-75^\circ \pm 30^\circ$. In both models, the inferred
source size increases with the total $1.3$ mm flux due to the emission
region becoming optically thick. The inferred sizes
($\text{DJ1}_{\text{FWHM}}$ and $\text{J2}_{\text{FWHM}}$) can be
expressed as power laws in flux as,

\begin{eqnarray}\label{sizes}
\text{DJ1}_{\text{FWHM}} &\simeq& (33\pm6)
\left(\frac{F_{\nu=230\text{GHz}}}{1\hspace{2pt}\text{Jy}}\right)^{.5} \mu\text{as}, \\
\text{J2}_{\text{FWHM}} &\simeq& (39\pm8)
\left(\frac{F_{\nu=230\text{GHz}}}{1\hspace{2pt}\text{Jy}}\right)^{.25} \mu\text{as},
\end{eqnarray}

\noindent where the uncertainties are the 1$\sigma$ scatter both from
the time variability and sky orientation. For the observed total flux
of $1.77$Jy found by \citet{tanetal2008}, the predicted sizes are both
$44\mu$as. VLBI observations of Sgr A* have found smaller
fluxes on micro-arcsecond scales than those from single dish
observations, in which case the values of $33\mu$as and $39\mu$as for
$1$Jy may be more appropriate for M87. 

\begin{figure}
\includegraphics[scale=0.55]{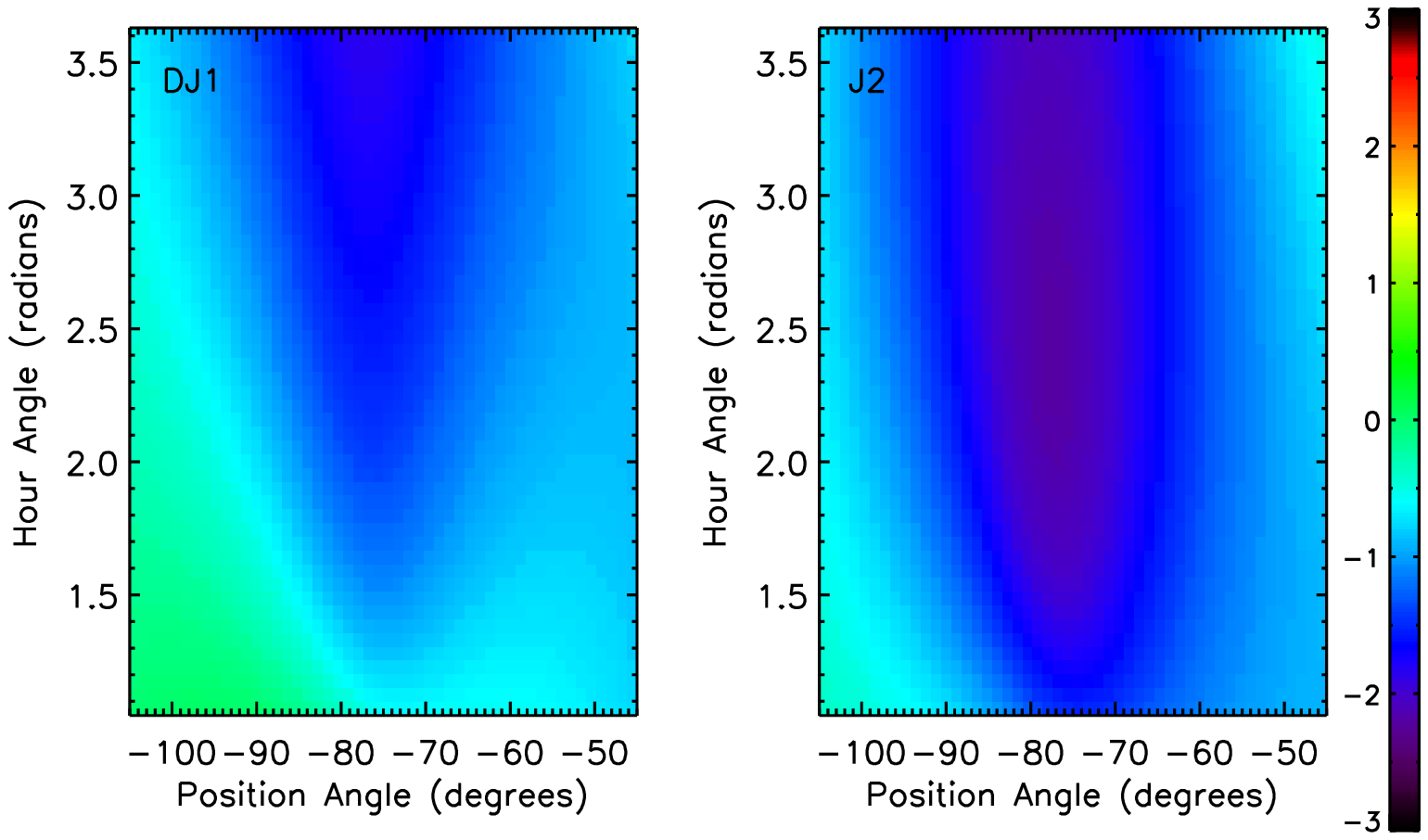}
\caption{\label{m87cphase}Predicted closure phase from the two fiducial models
  (left and right panels) at $1.3$ mm as functions of the position
  angle of the black hole spin axis and the projection of the
  Hawaii/California/Arizona baseline triangle on M87. For a large
  fraction of parameter space, the predicted closure phase differs
  significantly from zero, indicative of the asymmetric structure of
  the images.}
\end{figure}

We can also interpolate the visibilities from the fiducial models to the
baselines probed by telescopes to be added to the
mm-VLBI array in the near future in Chile\footnote{Atacama
  Pathfinder Experiment, APEX; Atacama Submillimeter Telescope
  Experiment, ASTE; Atacama Large Millimeter Array, ALMA.}
and Mexico,\footnote{Large Millimeter Telescope, LMT.} and the results
are shown in the bottom 
panel of Figure \ref{m87interpvis}. In both cases, the black hole
shadow is accessible to observations between Mexico and
Hawaii.\footnote{JCMT or Submillimeter Array, SMA.} In the
jet-dominated model, the shadow also appears on baselines between
Chile and Mexico. Finally, the VLBI closure phase provides an additional constraint for
any triangle of baselines, independent of that given by the visibility
amplitude. Predicted closure phases for our two fiducial models as functions of
position angle and the projection of the Hawaii/California/Arizona
baseline triangle on M87 are shown in Figure \ref{m87cphase}. The
predicted closure phase can deviate significantly from zero,
indicative of asymmetric structure, especially in the jet-dominated
model (J2). These results hold over the range of sky
orientations and simulation time considered.

The maximum brightness temperatures, $kT_b=I_\nu c^2/2\nu^2$
 in cgs units, where $I_\nu$ is the observed specific intensity, from
 the models are $3\times10^{10}$ K (DJ1) and $7\times10^{10}$ K
 (J2). The brightness temperature tracks the image intensity, and its
 maximum is larger for a more compact image. In the disc model, the
 maximum brightness temperature is in good agreement with the typical 
 temperature of the electrons producing the millimetre emission.

\section{Variability}
\label{sec:variability}
\begin{figure}
\includegraphics[scale=0.65]{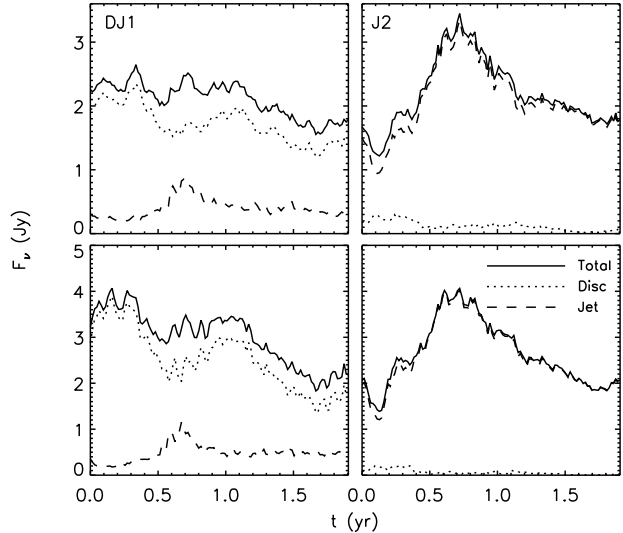}
\caption{\label{m87lcurves}Light curves from the two fiducial models
  (left and right panels) at $1.3$ mm (top row) and $0.87$ mm (bottom
  row). The disc variability is driven by MRI turbulence, as
  previously found in similar models of Sgr A*
  \citep{dexter2009,dexteretal2010}.}
\end{figure}

As discussed above, these radiative disc/jet models of M87 are also
time-dependent. Light curves from the fiducial 
models are shown in Figure \ref{m87lcurves} at $1.3$ mm and $0.87$ mm. The
variability in the jet model is from one extended event, while the disc model
light curve looks similar to that for the same simulation in the Sgr
A* modeling \citep[lower right panel of Figure 6 from
][]{dexteretal2010}. This disc variability is caused by fluctuations
in the particle density and magnetic field strength driven by MRI turbulence, and are
strongly correlated with the accretion rate. Both models are consistent with the
finding that M87 varies at about
$1\hspace{2pt}\text{Jy}\hspace{2pt}\text{yr}^{-1}$
\citep{steppeetal1988}. However, longer light curves would be required
to make a statistically significant statement about the
variability. This is particularly true in the jet model, where the
variability is the result of one event. It is
unclear whether this is due to a transient effect in the jet
formation/propagation or recurrent variable activity. 

\section{Discussion}
\label{sec:discussion}
We have created the first radiative disc/jet models of M87 based on GRMHD
simulations. The jet is formed, collimated and accelerated
self-consistently from an accretion flow within the ideal MHD approximation and the limitations of
the numerical scheme. The disc/jet boundary is taken as a contour in the ratio
of magnetic to rest mass energy density or in the particle specific enthalpy
as measured at infinity, with similar results in either case. The disc
portion is modeled as thermal electrons with a constant ion-electron
temperature ratio as in previous models of Sgr A*, while the
internal energy of the power law electrons in the jet region is scaled as a fixed fraction of the
magnetic energy density. The emission mechanism is synchrotron
radiation in both cases. Two separate types of models can describe the
high resolution spectrum of M87. In one class of models, the disc emission is negligible at all
frequencies and the jet produces the entire spectrum. In the other,
the jet produces the low and high frequency spectrum, while the
millimetre emission is produced by thermal disc electrons. 

\subsection{Favored Parameter Values}

The favored jet parameter combinations are
$\gamma_{\text{min}}=30-100$, $p=3.25-3.5$ and $\eta=0.02-0.10$. The
favored average accretion rate is $\dot{m}=(1-2) \times 10^{-4}$, or
$\dot{M}=(1-2) \times 10^{-3} 
M_\odot \text{yr}^{-1}$. For the jet portion, the parameters $\dot{m}$,
$\eta$ and $\gamma_{\text{min}}$ are partially degenerate. Ignoring
important effects of optical depth and the cutoff frequency on the
spectrum, $F_{\nu_{\text{max}}} \sim \eta 
\dot{M}^2 \gamma_{\text{min}}^4$, while $\nu_{\text{max}} \propto
\sqrt{\dot{m}} \gamma_{\text{min}}^2$. Jet spectra with
$\gamma_{\text{min}}=1000$ peak in the near-IR or optical,
inconsistent with VLBI observations of extended jet emission in the
radio. Smaller $\dot{m}$ 
is required at larger $\gamma_{\text{min}}$ to keep the spectral peak in the
millimetre. The fraction of magnetic energy density in non-thermal electrons, $\eta$, can
be used to scale to the correct 
normalization for any $\gamma_{\text{min}}$ and $\dot{m}$ as long as
it remains less than unity and small enough so that the jet remains
magnetically dominated (Equation \ref{eq:eta}). The
spectral slope, $p$, is fixed by the near IR and optical observations,
while the ion-electron temperature ratio $T_i/T_e$ determines whether
the disc or jet produces the emission at millimetre wavelengths. The
favored value for this simulation from Sgr A* models, $T_i/T_e=3$
\citep{dexteretal2010}, leads to a disc-dominated millimetre image. 

\subsection{Radiative Efficiency}

%This is complicated by the uncertain coupling between
%electrons and ions. Unless $T_i=T_e$, simulations with optically
%thin cooling \citep{fragile2009,moscibrodzkaetal2011} 

By ignoring radiation in the simulations, we have assumed that the
accretion flow is radiatively inefficient. We can check this
assumption by calculating the radiative efficiency, 
$\epsilon=L/\dot{M}c^2$, where $L \sim 2 \times
10^{42}\hspace{2pt}\text{erg}\hspace{2pt} 
\text{s}^{-1}$ is the bolometric luminosity of the models. This gives
$\epsilon \sim (1-4) \times 10^{-2}$, so that the radiative models are
radiatively inefficient and marginally self-consistent. The effects of
radiative cooling on the dynamical solution should be considered in future
simulations. Using axisymmetric GRMHD simulations,
\citet{moscibrodzkaetal2011} found lower accretion 
rates, $\dot{M} \simeq 10^{-4} M_{\odot} \text{yr}^{-1}$, and larger
bolometric luminosities, $L \sim
10^{44}\hspace{2pt}\text{erg}\hspace{2pt}\text{s}^{-1}$, 
leading to unphysically large radiative efficiencies ($\simeq 16$, see 
their Table 1). Their models used a smaller black hole mass and
$T_i=T_e$, which together could
 explain the small accretion rates found. The
 bolometric synchrotron luminosity $L \sim n B^2 T_e^2 M^3 \propto 
\dot{M}^2T_e^2/M^3$. Using the factor $\simeq 2.1$ smaller black hole
mass and factor of $2$ larger electron temperature 
then requires an order of magnitude decrease in accretion rate to
produce the observed flux, which explains the difference. Their models
also include Compton scattering, which is neglected here. Extending
our non-thermal jet spectrum to the X-ray does not significantly
change its bolometric luminosity, but including Compton
scattering would both increase the bolometric luminosity and allow us 
to include a constraint from the observed X-ray luminosity
\citep{wilsonyang2002}. In \citet{moscibrodzkaetal2011}, the spectral 
energy distribution from the simulation without radiative cooling
peaks at $\nu \gtrsim 10^{20}$Hz. It is unclear whether Compton
scattering would have such an important effect if included in our 3D
models. 

We can crudely estimate the relative significance of Compton
scattering for both models. In the jet model, 
the ratio of peak $\nu F_\nu$ for synchrotron and
self-Compton components in a homogeneous blob with a
power law distribution of electrons, ignoring all relativistic
and viewing effects is \citep[e.g.,][]{sariesin2001}:

\begin{equation}
  \frac{\nu F_\nu (\text{IC})}{\nu F_\nu (\text{Syn})}
  \sim \frac{2}{3}
  \frac{p-1}{p+3} \sigma_T \hspace{2pt}n\hspace{2pt} R\hspace{2pt} \gamma_{\text{min}}^{p-1} \hspace{2pt}\gamma_{\text{max}}^{3-p},
\end{equation}

\noindent where $\sigma_T$ is the Thompson cross-section and $R$ is
the size of the emitting blob. Scaling to reasonable parameters for
our jet model gives,

\begin{equation}
  \frac{\nu F_\nu (\text{IC})}{\nu F_\nu (\text{Syn})} \sim 0.5
  \left(\frac{R}{5M}\right) \left(\frac{n}{10^7 \hspace{2pt} \text{cm}^{-3}}\right) \left(\frac{\gamma_{\text{min}}}{50}\right)^{5/2} \left(\frac{\gamma_{\text{max}}}{10^5}\right)^{-1/2}.
\end{equation}

\noindent This estimate is not self-consistent. 
We have scaled the electron number density to the typical value for thermal disc
electrons which provide the scattering optical depth, but then assumed 
they have a power law energy distribution. This is a
conservative approach, since thermal electrons would have a much
sharper high energy cutoff. 

For the disc model, we use the approximate local prescription given in
\citet{esinetal1996} to calculate the Comptonization enhancement to
the bolometric thermal synchrotron luminosity. This enhancement
depends sensitively on the scattering optical depth and weakly on the
seed photon energy. The
scattering optical depth is taken to be $\tau = n \sigma H$, where
$\sigma$ is the Thomson scattering cross section and $H \equiv (H/R)
R$ is the disc height, where $H/R \simeq 0.2$ for the simulation
considered here. Comptonization is estimated from the peak frequency
of the synchrotron spectrum, $\nu_p \sim 10^{11-12}$Hz for the disc
models. The estimate is mostly independent of $\nu_p$ within this
range. Using typical values of $n=10^7 \hspace{2pt} \text{cm}^{-3}$, $R=1$ M, and
$T=2\times10^{10}$ K, we find the relative 
Compton luminosity $L_{\text{IC}}/L_{\text{Syn}}
\simeq 1$. Numerically integrating the Compton cooling rate over the
simulation domain gives a similar answer. However, this enhancement
factor is extremely sensitive to the assumed scattering optical
depth. For example, replacing the disc height $H$ with radius $R$ in
the expression above gives $L_C/L_S \sim 10$, or a radiative
efficiency $\epsilon \lesssim 1$, invalidating our assumption that
radiation can be added after the fact without changing the (thermo)dynamics of
the accretion flow.

From both estimates, it is clear that Compton
scattering is an important mechanism and will significantly affect the 
observed X-ray emission. However, we estimate that it may only lead to an
order unity correction in the bolometric luminosity of both of our
models. Including optically thin synchrotron and Compton cooling
self-consistently is an important goal for future simulations of M87,
and may significantly affect the resulting radiative
model. However, this is complicated by the uncertain ion-electron
coupling. Including optically thin radiative cooling in standard
single fluid simulations assumes perfect coupling between the ions and
electrons, since both species are cooled equally.

For completeness, we can also estimate the bolometric luminosity from
bremsstrahlung:

\begin{equation}
  L_{B} \sim 10^{40} \left(\frac{n}{10^7\hspace{2pt}\text{cm}^{-3}}\right)^2
  \left(\frac{T}{10^{10}\hspace{2pt}\text{K}}\right)^{1/2}
  \left(\frac{R}{10M}\right)^3 \text{erg} \hspace{2pt}\text{s}^{-1},
\end{equation}

\noindent a few orders of magnitude less than the synchrotron
luminosity. Including the large-scale accretion flow, with much
smaller particle densities and temperatures but a much larger volume,
will increase this estimate and both Compton scattering and
bremsstrahlung could be important emission mechanisms at X-ray energies.

\subsection{Images}
\label{sec:images}
Assuming that the M87 jet propagates along the black hole spin axis,
its orientation angle is constrained to be roughly $-75 \pm 30^\circ$
measured E of N. The inclination has been estimated to be
$\simeq25^\circ$ from the Lorentz factor of the jet
\citep{heinzbegelman1997,birettaetal1999}. For these parameters, we
can make predictions for ongoing mm-VLBI observations. On current 
baselines, we predict that M87 should appear as a compact source
similar to Sgr A*. The fiducial models have FWHM sizes of
$22-68$$\mu$as when fit with symmetric Gaussian models over the entire 
range of sky orientations and simulation time steps considered. For 
reasonable values of the total $1.3$ mm flux, the sizes are largely in
the range $33-44\mu$as, similar to the 
size of $37\mu$as found in the first mm-VLBI observations of Sgr A*
\citep{doeleman2008}.

The black hole shadow is accessible to future observations on baselines between 
Mexico and Hawaii and possibly Mexico and Chile. Although the
predicted Gaussian sizes for current telescopes are
nearly identical for both fiducial models, future observations with
additional telescopes or epochs 
should be able to distinguish between the two. This is because the 2D
structure of the images is significantly different and future
baselines will probe orientations where the predictions differ 
substantially. Future mm-VLBI
observations will also measure polarized emission on event horizon
scales, and including polarization in the radiative transfer
calculations is a goal for future work.

%Both images are crescents, as in the case of Sgr A*,
%with the relative importance of Doppler beaming and gravitational
%lensing set by the inclination and velocity profile of the
%material. 

Both images are crescents, as in the case of Sgr A*. When the emission
region is compact enough and the  
velocities are roughly Keplerian disc motion and helical jet motion, 
the resulting image is a crescent, qualitatively independent of the details of the
physics in the innermost part of the accretion flow. The sky
orientation (position angle) determines the orientations measured by
VLBI. Thus, our predictions for mm-VLBI with current and future telescopes
are fairly robust as long as the assumed geometry is 
accurate and the millimetre emission region is compact ($r \lesssim
10$ M), despite the fact that the fiducial models chosen here are 
only representative of a range of possibilities for 
explaining the spectral properties of M87. The compactness requirement
for the emission region is equivalent to assuming a steeply declining
emissivity with radius, and a small optical depth ($\tau \lesssim 1$)
to the emission region.

At inclinations of $10^\circ$ ($40^\circ$), the images become less
(more) Gaussian and more (less) circular. The crescent morphology and
size predictions are less sensitive to changes in inclination within
this range than to uncertainties in the model (DJ1 or J2) or position
angle.

We have assumed the new mass estimate of $6.4\times10^9 M_\odot$
\citep{gebhardtthomas2009,gebhardtetal2011} throughout. Changing the
black hole mass changes the favored parameter values, but does not
substantively change the types of feasible models or other major
results. The exception is in the size predictions for mm-VLBI, which
essentially scale with mass. If the black hole mass in M87 is
$3\times10^9 M_\odot$ \citep{marconietal1997}, then the predicted
sizes would be a factor $\simeq 2$ smaller. This means that mm-VLBI
can provide a model-dependent test of the black hole mass -- if a size
smaller than $\simeq22\mu$as is found, the smallest for our models over the
range of simulation time and valid position angles, it may indicate a smaller black hole
mass. Conversely, a relatively large measured size would favor the
larger black hole mass estimate, although it could also mean that the
emission at $1.3$ mm is more extended than found from current GRMHD
simulations. 

\subsection{Uncertainties}

There are many uncertainties in this analysis. The assumption that the
internal energy in non-thermal particles scales with magnetic field
energy density may be reasonable \citep[e.g., ][]{broderickmckinney2010},
but it is made out of necessity. Both the internal energy (pressure)
and mass density from the simulation are set by the artificial
numerical floor required for code stability when $b^2/\rho c^2 \gg 1$, the region of
interest for jet launching. Radiation is added in post-processing,
despite the fact that the models are found to be marginally
radiatively efficient. Both synchrotron and Compton scattering have
been found to be dynamically important in axisymmetric simulations of
M87 \citep{moscibrodzkaetal2011} with similar parameters to our disc/jet
model (DJ1). It will be possible 
to include those forms of cooling in future simulations, but a method
for evolving the non-thermal particle density self-consistently is much
more difficult. 

Previous semi-analytic disc/jet models of M87 have invoked a truncated
disc, with a low, constant particle density and magnetic field
strength throughout the inner disc. This configuration can produce the
observed radio emission. We find that the magnetic field strength in 
MBD and other GRMHD simulations falls off with radius ($\propto
r^{-1.2}$ for MBD). Producing the
observed flux then requires that the
disc emission peaks at millimetre wavelengths. Although none of our
models can explain the observed radio emission, the agreement is much better
if the models are fit to only the unresolved emission from the core of
radio images \citep[cf.][]{reynoldsetal1996,dimatteoetal2003}.

Both the disc and jet models used here are unlikely to be valid 
outside the innermost radii. The accretion flow solution is probably reliable only out to $r \simeq
10$ M, but the disc emission region is so compact that this does 
not affect the results. However, there are so
few studies of jets launched from GRMHD simulations that it is unclear
what the domain of validity is. This situation will improve in the
near future, as numerical techniques for propagating large-scale jets
improve and simulations are able to be evolved for longer physical
times due to the increase in available computational resources. The jet solution in 
a new simulation similar to MBD but with much higher resolution
(272$\times$128$\times$256), a larger radial boundary ($26000$ M) and a longer
duration ($20000$ M) is nearly identical to that of MBD within $r
\lesssim 8$ M ($B \propto r^{-1}$), but has a nearly flat radial magnetic field strength
profile outside of that. Preliminary images based on a single time
step from this simulation display extended (mas) structure in the radio, but
give nearly identical predictions for mm-VLBI of M87,
indicating that the low resolution used in MBD is unlikely to invalidate the
results of this paper. Models from this and other high resolution,
geometrically thick simulations \citep{mckinneyetal2012} will
be considered in a subsequent publication. 

\section{Summary}
\label{sec:summary}
We have constructed the first radiative disc/jet models of M87 based on a
general relativistic MHD simulation of a black hole accretion
flow. The main results of this study are: 

\begin{enumerate}

\item The spectral energy distribution of the core of M87
  can be explained with jet or disc/jet models.

\item In both types of models the images at $1.3$ mm are crescents from
  the combination of gravitational lensing and Doppler beaming of the
  compact emission region. 

\item The jet emission is produced in the \emph{counter-jet} near the
  pole in the immediate vicinity of the black hole ($r \lesssim 4$ M), while the disc emission
  is produced near the midplane in the inner radii $r \simeq 5$ M as
  found in previous models of Sgr A*.

\item For the favored viewing geometry of M87 based on analyses of the
  large-scale jet, we predict a Gaussian source structure with a size
  of $33-44\mu$as ($\simeq5$ Schwarzschild radii) for observations
  with current mm-VLBI 
  telescopes. The inferred size should increase slowly with increasing
  flux. The black hole shadow, direct evidence for an event
  horizon, may be detected on future baselines 
  between Hawaii (JCMT) and Chile (ALMA/APEX/ASTE).

\item The two types of models can be distinguished with mm-VLBI observations
  on baselines including telescopes in Mexico (LMT) and Chile
  (ALMA/APEX/ASTE), or by measuring the change in size between epochs of different total
  $1.3$ mm flux.

\end{enumerate}

\section*{acknowledgements}
We thank Shep Doeleman, Vincent Fish, and the anonymous referee for 
useful comments. This work was partially supported by NASA Earth \&
Space Science Fellowship NNX08AX59H (JD), STScI grant
HST-GO-11732.02-A, and NASA Chandra 
Fellowship PF7-80048 (JCM). 

\bibliographystyle{mn2e}
\bibliography{master}

\begin{thebibliography}{56}
\expandafter\ifx\csname natexlab\endcsname\relax\def\natexlab#1{#1}\fi

\bibitem[{{Baath} {et~al}\mbox{.}(1992){Baath}, {Rogers}, {Inoue}, {Padin},
  {Wright}, {Zensus}, {Kus}, {Backer}, {Booth}, {Carlstrom}, {Dickman},
  {Emerson}, {Hirabayashi}, {Hodges}, {Kobayashi}, {Lamb}, {Moran}, {Morimoto},
  {Plambeck}, {Predmore}, {Ronnang}, \& {Woody}}]{baathetal1992}
{Baath} L.~B. {et~al.}, 1992, \aap, 257, 31

\bibitem[{{Bardeen}(1973)}]{bardeen1973}
{Bardeen} J.~M., 1973, in Black holes (Les astres occlus), {DeWitt} B.~S.,
  {DeWitt} C., eds., New York: Gordon and Breach, p. 215

\bibitem[{{Biretta}, {Sparks} \& {Macchetto}(1999){Biretta}, {Sparks}, \&
  {Macchetto}}]{birettaetal1999}
{Biretta} J.~A., {Sparks} W.~B., {Macchetto} F., 1999, \apj, 520, 621

\bibitem[{{Broderick}(2006)}]{broderick2006}
{Broderick} A.~E., 2006, \mnras, 366, L10

\bibitem[{{Broderick} {et~al}\mbox{.}(2011){Broderick}, {Fish}, {Doeleman}, \&
  {Loeb}}]{brodericketal2011}
{Broderick} A.~E., {Fish} V.~L., {Doeleman} S.~S., {Loeb} A., 2011, \apj, 735,
  110

\bibitem[{{Broderick} \& {Loeb}(2009)}]{broderickloeb2009}
{Broderick} A.~E., {Loeb} A., 2009, \apj, 697, 1164

\bibitem[{{Broderick} \& {McKinney}(2010)}]{broderickmckinney2010}
{Broderick} A.~E., {McKinney} J.~C., 2010, \apj, 725, 750

\bibitem[{{De Villiers} {et~al}\mbox{.}(2005){De Villiers}, {Hawley}, {Krolik},
  \& {Hirose}}]{devilliersetal2005}
{De Villiers} J.-P., {Hawley} J.~F., {Krolik} J.~H., {Hirose} S., 2005, \apj,
  620, 878

\bibitem[{{Dexter}(2011)}]{dexter2011}
{Dexter} J., 2011, PhD thesis, University of Washington

\bibitem[{{Dexter} \& {Agol}(2009)}]{dexteragol2009}
{Dexter} J., {Agol} E., 2009, \apj, 696, 1616

\bibitem[{{Dexter}, {Agol} \& {Fragile}(2009){Dexter}, {Agol}, \&
  {Fragile}}]{dexter2009}
{Dexter} J., {Agol} E., {Fragile} P.~C., 2009, \apjl, 703, L142

\bibitem[{{Dexter} {et~al}\mbox{.}(2010){Dexter}, {Agol}, {Fragile}, \&
  {McKinney}}]{dexteretal2010}
{Dexter} J., {Agol} E., {Fragile} P.~C., {McKinney} J.~C., 2010, \apj, 717,
  1092

\bibitem[{{Di Matteo} {et~al}\mbox{.}(2003){Di Matteo}, {Allen}, {Fabian},
  {Wilson}, \& {Young}}]{dimatteoetal2003}
{Di Matteo} T., {Allen} S.~W., {Fabian} A.~C., {Wilson} A.~S., {Young} A.~J.,
  2003, \apj, 582, 133

\bibitem[{{Doeleman} {et~al}\mbox{.}(2008){Doeleman}, {Weintroub}, {Rogers},
  {Plambeck}, {Freund}, {Tilanus}, {Friberg}, {Ziurys}, {Moran}, {Corey},
  {Young}, {Smythe}, {Titus}, {Marrone}, {Cappallo}, {Bock}, {Bower},
  {Chamberlin}, {Davis}, {Krichbaum}, {Lamb}, {Maness}, {Niell}, {Roy},
  {Strittmatter}, {Werthimer}, {Whitney}, \& {Woody}}]{doeleman2008}
{Doeleman} S.~S. {et~al.}, 2008, \nat, 455, 78

\bibitem[{{Esin} {et~al}\mbox{.}(1996){Esin}, {Narayan}, {Ostriker}, \&
  {Yi}}]{esinetal1996}
{Esin} A.~A., {Narayan} R., {Ostriker} E., {Yi} I., 1996, \apj, 465, 312

\bibitem[{{Falcke}, {Melia} \& {Agol}(2000){Falcke}, {Melia}, \&
  {Agol}}]{falcke}
{Falcke} H., {Melia} F., {Agol} E., 2000, \apjl, 528, L13

\bibitem[{{Fish} {et~al}\mbox{.}(2011){Fish}, {Doeleman}, {Beaudoin},
  {Blundell}, {Bolin}, {Bower}, {Chamberlin}, {Freund}, {Friberg}, {Gurwell},
  {Honma}, {Inoue}, {Krichbaum}, {Lamb}, {Marrone}, {Moran}, {Oyama},
  {Plambeck}, {Primiani}, {Rogers}, {Smythe}, {SooHoo}, {Strittmatter},
  {Tilanus}, {Titus}, {Weintroub}, {Wright}, {Woody}, {Young}, \&
  {Ziurys}}]{fishetal2011}
{Fish} V.~L. {et~al.}, 2011, \apjl, 727, L36+

\bibitem[{{Fragile} {et~al}\mbox{.}(2007){Fragile}, {Blaes}, {Anninos}, \&
  {Salmonson}}]{fragile2007}
{Fragile} P.~C., {Blaes} O.~M., {Anninos} P., {Salmonson} J.~D., 2007, \apj,
  668, 417

\bibitem[{{Fuerst} \& {Wu}(2004)}]{fuerstwu2004}
{Fuerst} S.~V., {Wu} K., 2004, \aap, 424, 733

\bibitem[{{Gammie}, {McKinney} \& {T{\'o}th}(2003){Gammie}, {McKinney}, \&
  {T{\'o}th}}]{gammie2003}
{Gammie} C.~F., {McKinney} J.~C., {T{\'o}th} G., 2003, \apj, 589, 444

\bibitem[{{Gebhardt} {et~al}\mbox{.}(2011){Gebhardt}, {Adams}, {Richstone},
  {Lauer}, {Faber}, {G{\"u}ltekin}, {Murphy}, \& {Tremaine}}]{gebhardtetal2011}
{Gebhardt} K., {Adams} J., {Richstone} D., {Lauer} T.~R., {Faber} S.~M.,
  {G{\"u}ltekin} K., {Murphy} J., {Tremaine} S., 2011, \apj, 729, 119

\bibitem[{{Gebhardt} \& {Thomas}(2009)}]{gebhardtthomas2009}
{Gebhardt} K., {Thomas} J., 2009, \apj, 700, 1690

\bibitem[{{Goldston}, {Quataert} \& {Igumenshchev}(2005){Goldston}, {Quataert},
  \& {Igumenshchev}}]{goldston2005}
{Goldston} J.~E., {Quataert} E., {Igumenshchev} I.~V., 2005, \apj, 621, 785

\bibitem[{{Gracia} {et~al}\mbox{.}(2009){Gracia}, {Vlahakis}, {Agudo},
  {Tsinganos}, \& {Bogovalov}}]{graciaetal2009}
{Gracia} J., {Vlahakis} N., {Agudo} I., {Tsinganos} K., {Bogovalov} S.~V.,
  2009, \apj, 695, 503

\bibitem[{{Hada} {et~al}\mbox{.}(2011){Hada}, {Doi}, {Kino}, {Nagai},
  {Hagiwara}, \& {Kawaguchi}}]{hadaetal2011}
{Hada} K., {Doi} A., {Kino} M., {Nagai} H., {Hagiwara} Y., {Kawaguchi} N.,
  2011, \nat, 477, 185

\bibitem[{{Heinz} \& {Begelman}(1997)}]{heinzbegelman1997}
{Heinz} S., {Begelman} M.~C., 1997, \apj, 490, 653

\bibitem[{{Hilburn} \& {Liang}(2011)}]{hilburnliang2011}
{Hilburn} G., {Liang} E., 2011, ArXiv e-prints

\bibitem[{{Junor}, {Biretta} \& {Livio}(1999){Junor}, {Biretta}, \&
  {Livio}}]{junoretal1999}
{Junor} W., {Biretta} J.~A., {Livio} M., 1999, \nat, 401, 891

\bibitem[{{Legg} \& {Westfold}(1968)}]{leggwestfold1968}
{Legg} M.~P.~C., {Westfold} K.~C., 1968, \apj, 154, 499

\bibitem[{{Leung}, {Gammie} \& {Noble}(2011){Leung}, {Gammie}, \&
  {Noble}}]{leungetal2011}
{Leung} P.~K., {Gammie} C.~F., {Noble} S.~C., 2011, \apj, 737, 21

\bibitem[{{Ly}, {Walker} \& {Wrobel}(2004){Ly}, {Walker}, \&
  {Wrobel}}]{lyetal2004}
{Ly} C., {Walker} R.~C., {Wrobel} J.~M., 2004, \aj, 127, 119

\bibitem[{{Marconi} {et~al}\mbox{.}(1997){Marconi}, {Axon}, {Macchetto},
  {Capetti}, {Sparks}, \& {Crane}}]{marconietal1997}
{Marconi} A., {Axon} D.~J., {Macchetto} F.~D., {Capetti} A., {Sparks} W.~B.,
  {Crane} P., 1997, \mnras, 289, L21

\bibitem[{{McKinney}(2006{\natexlab{a}})}]{mckinney2006ff}
{McKinney} J.~C., 2006{\natexlab{a}}, \mnras, 367, 1797

\bibitem[{{McKinney}(2006{\natexlab{b}})}]{mckinney2006}
---, 2006{\natexlab{b}}, \mnras, 368, 1561

\bibitem[{{McKinney} \& {Blandford}(2009)}]{mckinneyblandford2009}
{McKinney} J.~C., {Blandford} R.~D., 2009, \mnras, 394, L126

\bibitem[{{McKinney}, {Tchekhovskoy} \& {Blandford}(2012){McKinney},
  {Tchekhovskoy}, \& {Blandford}}]{mckinneyetal2012}
{McKinney} J.~C., {Tchekhovskoy} A., {Blandford} R.~D., 2012, ArXiv e-prints

\bibitem[{{Mo{\'s}cibrodzka} {et~al}\mbox{.}(2011){Mo{\'s}cibrodzka}, {Gammie},
  {Dolence}, \& {Shiokawa}}]{moscibrodzkaetal2011}
{Mo{\'s}cibrodzka} M., {Gammie} C.~F., {Dolence} J.~C., {Shiokawa} H., 2011,
  \apj, 735, 9

\bibitem[{{Mo{\'s}cibrodzka} {et~al}\mbox{.}(2009){Mo{\'s}cibrodzka}, {Gammie},
  {Dolence}, {Shiokawa}, \& {Leung}}]{moscibrodzka2009}
{Mo{\'s}cibrodzka} M., {Gammie} C.~F., {Dolence} J.~C., {Shiokawa} H., {Leung}
  P.~K., 2009, \apj, 706, 497

\bibitem[{{Noble} {et~al}\mbox{.}(2006){Noble}, {Gammie}, {McKinney}, \& {Del
  Zanna}}]{noble2006}
{Noble} S.~C., {Gammie} C.~F., {McKinney} J.~C., {Del Zanna} L., 2006, \apj,
  641, 626

\bibitem[{{Pauliny-Toth} {et~al}\mbox{.}(1981){Pauliny-Toth}, {Preuss},
  {Witzel}, {Graham}, {Kellerman}, \& {Ronnang}}]{paulinytothetal1981}
{Pauliny-Toth} I.~I.~K., {Preuss} E., {Witzel} A., {Graham} D., {Kellerman}
  K.~I., {Ronnang} B., 1981, \aj, 86, 371

\bibitem[{{Perlman} {et~al}\mbox{.}(2001){Perlman}, {Biretta}, {Sparks},
  {Macchetto}, \& {Leahy}}]{perlmanetal2001}
{Perlman} E.~S., {Biretta} J.~A., {Sparks} W.~B., {Macchetto} F.~D., {Leahy}
  J.~P., 2001, \apj, 551, 206

\bibitem[{{Perlman} {et~al}\mbox{.}(2007){Perlman}, {Mason}, {Packham},
  {Levenson}, {Elitzur}, {Schaefer}, {Imanishi}, {Sparks}, \&
  {Radomski}}]{perlmanetal2007}
{Perlman} E.~S. {et~al.}, 2007, \apj, 663, 808

\bibitem[{{Reynolds} {et~al}\mbox{.}(1996){Reynolds}, {Di Matteo}, {Fabian},
  {Hwang}, \& {Canizares}}]{reynoldsetal1996}
{Reynolds} C.~S., {Di Matteo} T., {Fabian} A.~C., {Hwang} U., {Canizares}
  C.~R., 1996, \mnras, 283, L111

\bibitem[{{Sari} \& {Esin}(2001)}]{sariesin2001}
{Sari} R., {Esin} A.~A., 2001, \apj, 548, 787

\bibitem[{{Shcherbakov}, {Penna} \& {McKinney}(2010){Shcherbakov}, {Penna}, \&
  {McKinney}}]{shcherbakovetal2011}
{Shcherbakov} R.~V., {Penna} R.~F., {McKinney} J.~C., 2010, ArXiv e-prints

\bibitem[{{Sparks}, {Biretta} \& {Macchetto}(1996){Sparks}, {Biretta}, \&
  {Macchetto}}]{sparksetal1996}
{Sparks} W.~B., {Biretta} J.~A., {Macchetto} F., 1996, \apj, 473, 254

\bibitem[{{Spencer} \& {Junor}(1986)}]{spencerjunor1986}
{Spencer} R.~E., {Junor} W., 1986, \nat, 321, 753

\bibitem[{{Steppe} {et~al}\mbox{.}(1988){Steppe}, {Salter}, {Chini}, {Kreysa},
  {Brunswig}, \& {Lobato Perez}}]{steppeetal1988}
{Steppe} H., {Salter} C.~J., {Chini} R., {Kreysa} E., {Brunswig} W., {Lobato
  Perez} J., 1988, \aaps, 75, 317

\bibitem[{{Tan} {et~al}\mbox{.}(2008){Tan}, {Beuther}, {Walter}, \&
  {Blackman}}]{tanetal2008}
{Tan} J.~C., {Beuther} H., {Walter} F., {Blackman} E.~G., 2008, \apj, 689, 775

\bibitem[{{Tchekhovskoy}, {McKinney} \& {Narayan}(2007){Tchekhovskoy},
  {McKinney}, \& {Narayan}}]{tchekhovskoyetal2007}
{Tchekhovskoy} A., {McKinney} J.~C., {Narayan} R., 2007, \mnras, 379, 469

\bibitem[{{Walker} {et~al}\mbox{.}(2008){Walker}, {Ly}, {Junor}, \&
  {Hardee}}]{walkeretal2008}
{Walker} R.~C., {Ly} C., {Junor} W., {Hardee} P.~J., 2008, Journal of Physics
  Conference Series, 131, 012053

\bibitem[{{Westfold}(1959)}]{westfold1959}
{Westfold} K.~C., 1959, \apj, 130, 241

\bibitem[{{Wilson} \& {Yang}(2002)}]{wilsonyang2002}
{Wilson} A.~S., {Yang} Y., 2002, \apj, 568, 133

\bibitem[{{Yuan}(2000)}]{yuan2000}
{Yuan} F., 2000, \mnras, 319, 1178

\bibitem[{{Yuan}, {Quataert} \& {Narayan}(2003){Yuan}, {Quataert}, \&
  {Narayan}}]{yuanquataert2003}
{Yuan} F., {Quataert} E., {Narayan} R., 2003, \apj, 598, 301

\bibitem[{{Zakamska}, {Begelman} \& {Blandford}(2008){Zakamska}, {Begelman}, \&
  {Blandford}}]{zakamskaetal2008}
{Zakamska} N.~L., {Begelman} M.~C., {Blandford} R.~D., 2008, \apj, 679, 990

\end{thebibliography}

\label{lastpage}

\end{document}